\documentclass[12pt]{article}

	\usepackage[nohead,margin=1in]{geometry}	
	\usepackage{booktabs}
	\usepackage{tabularx}
    \newcolumntype{s}{>{\hsize=.25\hsize}X}
    \usepackage{longtable}
    \usepackage{afterpage}
	\usepackage{etoolbox}
	\usepackage{appendix}
	\usepackage{comment}
	\usepackage{tikz,pgfplots,pgfplotstable}
	\pgfplotsset{compat=1.17}
    \usetikzlibrary{decorations.pathreplacing}

	\usepackage{pdfpages}
	
		\usepackage{setspace}
				
		\def\notesize{\footnotesize}
		
		\newenvironment{tablenotes}[1][Note]{\begin{minipage}[t]{\linewidth}\notesize{\itshape#1: }}{\end{minipage}}
		
		\newenvironment{figurenotes}[1][Note]{\begin{minipage}[t]{\linewidth}\notesize{\itshape#1: }}{\end{minipage}}
		
		\def\@shortTitle{\@title}
		\def\shortTitle#1{\gdef\@shortTitle{#1}}
		
		\def\@draftSpacing{1}
		\def\draftSpacing#1{\gdef\@draftSpacing{#1}}
		
		\long\gdef\@pubYear{Year}
		\long\def\pubYear#1{\long\gdef\@pubYear{#1}}
		
		\long\gdef\@pubMonth{Month}
		\long\def\pubMonth#1{\long\gdef\@pubMonth{#1}}
		
		\long\gdef\@pubVolume{Volume}
		\long\def\pubVolume#1{\long\gdef\@pubVolume{#1}}
		
		\long\gdef\@pubIssue{Issue}
		\long\def\pubIssue#1{\long\gdef\@pubIssue{#1}}
			
		\long\gdef\@JEL{}
		\long\def\JEL#1{\long\gdef\@JEL{#1}}
		
		\long\gdef\@Keywords{}
		\long\def\Keywords#1{\long\gdef\@Keywords{#1}}

	\usepackage{amssymb}
	\usepackage{amsmath}

	\usepackage{graphicx}
	\usepackage{rotating}
	\usepackage{subcaption}
	
	\usepackage{xr-hyper}
	\usepackage{hyperref}
	\usepackage{footnotebackref}
	
	\usepackage{xcolor}
	\hypersetup{
	colorlinks,
	allcolors={blue!50!black},
	bookmarksnumbered
	}
	
	\usepackage[utf8x]{luainputenc}
	\PrerenderUnicode{é}
	\PrerenderUnicode{á}
	\PrerenderUnicode{ũ}

	\usepackage{csquotes}
	
	\usepackage{pdflscape}

	\usepackage[labelfont=bf,labelsep=newline,justification=centering]{caption}	
	
    \usepackage{bbm}

	\usepackage{natbib}			
	\usepackage{multibib}
		\newcites{appendix}{Appendix References}
	\usepackage{titling}
	\makeatletter
	\def\@makefnmark{
	 	\leavevmode
		\raise.9ex\hbox{\fontsize\sf@size\z@\normalfont\tiny\@thefnmark}}
	\makeatother	
	
	\usepackage[nolists, noheads, nomarkers, figuresfirst]{endfloat}
	
	\newenvironment{landscapefig}
	{\begin{landscape}
	}
	{
	\end{landscape}
	}
	\DeclareDelayedFloatFlavor{landscape}{table}
    \DeclareDelayedFloatFlavor{landscapefig}{figure}

\usepackage{tikz}
\usetikzlibrary{decorations.pathreplacing}
\usetikzlibrary{shapes.geometric, arrows}
\tikzstyle{startstop} = [rectangle, minimum width=4cm,text centered, draw=black]
\tikzstyle{startstop2} = [rectangle, minimum width=4cm,text centered, draw=white]
\tikzstyle{process} = [rectangle,text centered, draw=black]
\tikzstyle{decision} = [diamond, minimum width=3cm, minimum height=1cm, text centered, draw=black]
\tikzstyle{arrow} = [thick,->,>=stealth]

\begin{document}
\phantomsection
\addcontentsline{toc}{section}{Appointments: A More Effective Commitment Device for Health Behaviors}
\title{Appointments:\\A More Effective Commitment Device\\for Health Behaviors}


\author{Laura Derksen, Jason T. Kerwin, Natalia Ordaz Reynoso, and Olivier Sterck\thanks{Derksen: Department of Management, University of Toronto (\href{mailto:laura.derksen@rotman.utoronto.ca}{laura.derksen@rotman.utoronto.ca}); 
Kerwin: Department of Applied Economics, University of Minnesota, and J-PAL (\href{mailto:jkerwin@umn.edu}{jkerwin@umn.edu}); Ordaz Reynoso: AXA Research Lab on Gender Equality, Dondena Research Centre, Bocconi University (\href{mailto:natalia.ordaz@unibocconi.it}{natalia.ordaz@unibocconi.it}); Sterck: Department of International Development, University of Oxford (\href{mailto:olivier.sterck@qeh.ox.ac.uk}{olivier.sterck@qeh.ox.ac.uk}). We thank Susan Watkins for her extensive guidance and mentorship on this project. We are grateful for insightful comments from Manuela Angelucci, Liang Bai, Leah Bevis, Lasse Brune, Gharad Bryan, Eric Chyn, Jon de Quidt, Stefano DellaVigna, Kate Dovel, Jessica Gallant, Erick Gong, Anett John, Maggie McConnell, Rachael Meager, Mauricio Romero, and Simone Schaner, and from seminar participants at PacDev and from MPC, IAS, and the Applied Economics department at the University of Minnesota. This project would not have been possible without the hard work of our fantastic fieldwork supervisor, Abdul Chilungo, the efforts of our wonderful team of survey interviewers and HDAs, and the participation of the men who generously shared their time with us to take part in the study. We thank Dina O'Brien, Ethan Sansom, Tim White and Heather Wong for their excellent research assistance. Connaught, GATE and SRCHSS at the University of Toronto and CIFAP at the University of Minnesota provided grant funding for this research. Kerwin gratefully acknowledges support from the Minnesota Population Center (P2C HD041023) funded through a grant from the Eunice Kennedy Shriver National Institute for Child Health and Human Development (NICHD), and from a UMN Institute for Advanced Study Residential Fellowship. This study was reviewed and approved by IRBs in Malawi (National Health Sciences Research Committee, protocol \# 19/03/2268) and at the University of Minnesota (protocol \# STUDY00005587) and the University of Toronto (protocol \# 00037491), and is registered with the AEA RCT Registry under registration number AEARCTR-0004295. All errors and omissions are our own.}}

\date{October 8, 2021}

\maketitle

\begin{abstract}

Health behaviors are plagued by self-control problems, and commitment devices are frequently proposed as a solution. We show that a simple alternative works even better: appointments. We randomly offer HIV testing appointments and financial commitment devices to high-risk men in Malawi. Appointments are much more effective than financial commitment devices, more than doubling testing rates. In contrast, most men who take up financial commitment devices lose their investments. Appointments address procrastination without the potential drawback of commitment failure, and also address limited memory problems. Appointments have the potential to increase demand for healthcare in the developing world.

\bigskip


\end{abstract}

\setstretch{1.5}

\thispagestyle{empty}

\pagebreak
	
\clearpage

\setcounter{page}{1}

\phantomsection
\addcontentsline{toc}{section}{Introduction}

Health behaviors are plagued by commitment problems. The decision to seek healthcare, exercise more, or to eat a healthier diet requires people to incur short-term costs for long-term gains and to follow through with plans. This makes health behaviors particularly prone to behavioral biases (\citealt{kessler_behavioural_2015}), particularly self-control problems (\citealt{dellavigna_paying_2006}). One potential solution to these challenges is to use commitment devices (\citealt{laibson1997golden,gul2001temptation,halpernCommitmentContractsWay2012,laibson2015don}). Financial commitment devices can encourage healthier behaviors in settings ranging from smoking (\citealt{ginePutYourMoney2010}) and drinking (\citealt{schilbach_alcohol_2019}) to gym attendance (\citealt{royerIncentivesCommitmentsHabit2015}).\footnote{Commitment devices have also been extensively studied in non-health settings. For example, commitment devices can be an effective way to promote savings (\citealt{ashraf2006tying}), and increase work effort (\citealt{kaurSelfControlWork2015}). For a review of the literature on commitment devices, see \cite{bryan2010commitment}.} However, many people who demand financial commitment devices do not follow through (\citealt{john_when_2020}). In the average study of financial commitments tied to health behaviors, 65 percent of people lose the money they staked on their own behavior (\autoref{tab_literature}). Moreover, there is evidence that these commitment devices are actually welfare-diminishing (\citealt{bai_self-control_2020}).

This paper provides evidence that health appointments help address behavioral barriers, and that they do so more effectively than financial commitment devices. We find that appointments address the same procrastination and commitment problems that plague health behaviors, but with greater success and fewer downsides.

We compare appointments and financial commitment devices using a randomized field experiment on HIV testing. Our sample consists of 1,232 high-risk men recruited at urban bars in Zomba, Malawi. Increasing HIV testing is a major public health challenge in sub-Saharan Africa, as delays in treatment lead to unnecessary deaths and new infections.\footnote{Approximately 680,000 people died of AIDS in 2020, and 1.5 million became newly infected with HIV (\citealt{UNAIDS2020}). Antiretroviral therapy prevents both death and HIV transmission, but many people living with HIV remain undiagnosed.} Participants were randomized to four study arms: a control group, appointments, commitment devices, or appointments plus commitment devices. Men in either appointments arm could sign up for an HIV testing appointment at a clinic, date, and time of their choosing over the next two months. They received a phone call reminder two days before their appointment with the option to reschedule. The commitment device allowed men to stake USD \$1.38 (approximately 80 percent of daily GDP per capita), taken out of their study compensation, on showing up at any testing clinic.\footnote{Each participant received USD \$2.76 as study compensation, as well as a voucher worth USD \$0.69 if presented at any HIV testing clinic in the city. These clinics offer HIV testing for free on a walk-in basis. } We elicited incentivized preferences for this device for all study participants, implementing the choice for men who were assigned to one of the commitment device arms. 

Offering men appointments sharply increases the HIV testing rate, raising it by 16 percentage points.\footnote{All our inferences are robust to variations in the choice of controls and outcome variable, and to the \cite{benjamini2006adaptive} false discovery rate correction for multiple testing.} This is a 140 percent increase relative to the control-group testing rate of 11 percent. Using the randomized assignment as an instrument, we find that the appointments increase HIV testing among the two-thirds of men who sign up for them by 23 percentage points.

In contrast, commitment devices are far less effective, and backfire for a substantial fraction of the men who sign up for them. About 50 percent of study participants demand a commitment device, suggesting that men know that self-control issues are a barrier to HIV testing. But the treatment effect of the commitment devices on HIV testing is just 8 percentage points, half that of the appointments, and we can reject the equality of the two effects at the 0.05 level. Moreover, most of the men who signed up for a commitment device were made worse off: 59 percent of those who received a commitment device lost the money they staked on getting tested. This failure of commitment parallels findings by \cite{john_when_2020}, \cite{bai_self-control_2020}, and \cite{buehren2020limits}. Combining both appointments and commitment devices leads to an 18 percentage-point increase in HIV testing. This estimate is only slightly higher than, and not significantly different from, the impact of appointments alone, suggesting that appointments act as a superior substitute for financial commitment devices.

Appointments are a highly cost-effective way of increasing HIV testing. We find that the cost per additional HIV test induced by an appointment was USD \$2.69, as compared with USD \$3.01 for a commitment device.\footnote{In this calculation, we assume that the money for the commitment device is paid by the participant. These costs only include the cost of implementation by a health provider and not the cost to the participant. If commitment losses are included in the cost, the cost of the commitment device rises to \$19.61 per additional person tested. The cost of scheduling an appointment includes mobile phone credit costs and marginal labor costs but not the fixed cost of finding and initiating contact with participants.} Appointments thus compare favorably with cash payments, which cost \$11 per additional person who learns their test result (\citealt{thorntonDemandImpactLearning2008}).\footnote{Information campaigns, on the other hand, can have low marginal costs if delivered through mass media, and a well-designed intervention can increase HIV testing by 40 to 100 percent (\citealt{banerjee2019entertaining, derksen_love_2021}).} The high cost-effectiveness of appointments for promoting HIV testing is particularly policy relevant given the importance of early diagnosis and treatment in preventing both AIDS deaths (\citealt{insight_start_study_group_initiation_2015}) and the spread of HIV (\citealt{cohen_prevention_2011}). That these effects are for men is especially important: men in Malawi are less likely than women to seek treatment for HIV and more likely to die of AIDS (\citealt{dovel2015men}). 

The larger benefits and lower costs of appointments naturally raise the question of why they work so well. Appointments, as typically implemented, likely address several different behavioral biases simultaneously, in ways that cannot easily be separated into quantifiable or comparable additive effects. However, we find evidence suggesting that two mechanisms are particularly important drivers of the success of appointments.

One reason that appointments are such an effective substitute for financial commitments is that they may operate as social commitment devices. Men who fail to show up for their appointments waste the time of clinic staff, and their absences may be noted and commented on. Appointments substitute nearly perfectly for financial commitment devices: the combination of the two interventions works almost exactly as well as an appointment alone. Looking at the study arm that received only appointments, the appointments are far more effective for men who wanted a commitment device but did not receive one---and for the subset of men who want a commitment device, getting only an appointment is almost as good as getting an appointment plus the commitment device. These effects, which are much larger those typically estimated in studies that involve making a private plan (e.g. \citealt{macis_using_2021}), suggest that appointments create social pressure to follow through, helping people overcome the same self-control issues that create demand for a commitment device.\footnote{Our results are consistent with \cite{gneezy_fine_2000}, who find that social pressure is more effective than financial incentives at inducing people to pick up their children on time from daycare. Exploiting social pressure as a commitment device is also seen in financial markets; it is a key part of the design of traditional microcredit products (\citealt{de_aghion_economics_2005}), and a similar mechanism is utilized in \cite{karlan_borrow_2012} for debt reduction. Social pressure from peers can also affect health decisions (\citealt{karingSocialSignalingChildhood2018}).}

A second reason that appointments are such an effective behavioral intervention is that they come with reminders, which seem to help overcome limited memory issues (\citealt{ericson_forgetting_2011,haushofer_cost_2015,ericson_interaction_2017}). HIV testing in Malawi is typically walk-in only, and private, informal plans to get tested may be easily forgotten.\footnote{Consistent with men forgetting their plans to get tested, fewer than half of men who enrolled in a financial commitment device actually got an HIV test.} In our study, men with appointments received phone-call reminders two days beforehand. We find evidence consistent with forgetfulness, and show that reminders may help address this issue. Reminders are a well-established public health intervention, with proven benefits for outcomes ranging from attendance at regular medical visits (\citealt{gurolurganci_mobile_2013}, \citealt{altmannNudgesDentist2014}) to vaccination uptake (\citealt{banerjee_selecting_2021}) to adherence to medication (\citealt{vervloet_effectiveness_2012}).\footnote{Forgetting also leads to important failures to optimize in life insurance (\citealt{Gottlieb_Smetters_2021}) and retirement savings (\citealt{goodman_abandoned_2021}). Reminders can help people make better financial decisions (\citealt{karlan_getting_2016}), raise charitable giving rates (\citealt{Rogers_Milkman_2016}), and increase the uptake of tutoring (\citealt{pugatch_nudging_2018}). More broadly, the evidence of forgetting that we see in our study is consistent with models of limited attention; see \cite{gabaix_chapter_2019} for a review.} The time pattern of HIV testing suggests the appointment reminders were important: while testing in the control group tapers off to nothing within a couple of weeks of recruitment, men in the appointments arm continue to come in for testing late into the study period. Crucially, this pattern holds even for men who do not actually get tested on their appointment date. Although this evidence suggests reminders play an important role, they are not the only mechanism at play. \cite{salvadori2020appointment} find that reminders explain about a third of the increase in HIV testing induced by appointments. In our study, conditional on receiving a reminder, 67 percent of participants who get tested do so on their exact appointment date (two days after the reminder), and we observe a clear spike in testing on that date. If the impact of appointments were due to reminders alone, we would expect visits to occur throughout the days or weeks after the reminder, with no spike on the exact appointment date. This spike in testing could reflect the desire to honor the social commitment, to follow through on a personal plan, or to experience reduced wait times. This last possibility is unlikely in our context: qualitative data indicates that the HIV testing clinics in our sample (and across Malawi) operate below capacity and there is typically no wait.\footnote{Indeed, we might expect larger effect sizes, higher demand for appointments, and different logistical challenges in a setting where wait times are long (\citealt{hakimov_how_2021}).}

Our findings build on the literature on using commitment devices to promote health behaviors, by showing that appointments can achieve larger benefits than financial commitments, without the drawbacks. We compare the two interventions directly, using random assignment of participants from a single sample, in a particularly relevant context. In general, studies of commitment devices evaluate a single intervention, or compare different types of financial commitments; \autoref{tab_literature} summarizes the existing literature on commitment devices to encourage health behaviors.\footnote{\cite{schilbach_alcohol_2019} presents a similar table, focusing on take-up rates alone and covering other domains beyond health behaviors.} The effects we estimate for commitment devices are within the range of those found in previous studies. In contrast, appointments compare very favorably to other health commitment devices by every metric. Their take-up rate ranks in the top quintile. Unlike commitment devices, appointments do not cause large numbers of people to lose money due to a failure to follow through.\footnote{A missed appointment may carry costs in terms of shame (\citealt{butera_measuring_2021}), which we are unable to directly measure. However, failed commitments likely carry psychic costs as well.} Most strikingly, their treatment effect is four times as large as that of the average financial commitment device, and over 50 percent larger than the most successful intervention from the extant literature (the combination of a personalized financial commitment and a discount from \citealt{bai_self-control_2020}).

\begin{landscape}
\begin{table}[htbp!]
\begin{center}
 \caption{Take-up and Effects of Commitment Devices for Health Behaviors\label{tab_literature}} 
 \includegraphics[width=0.93\linewidth,page=1]{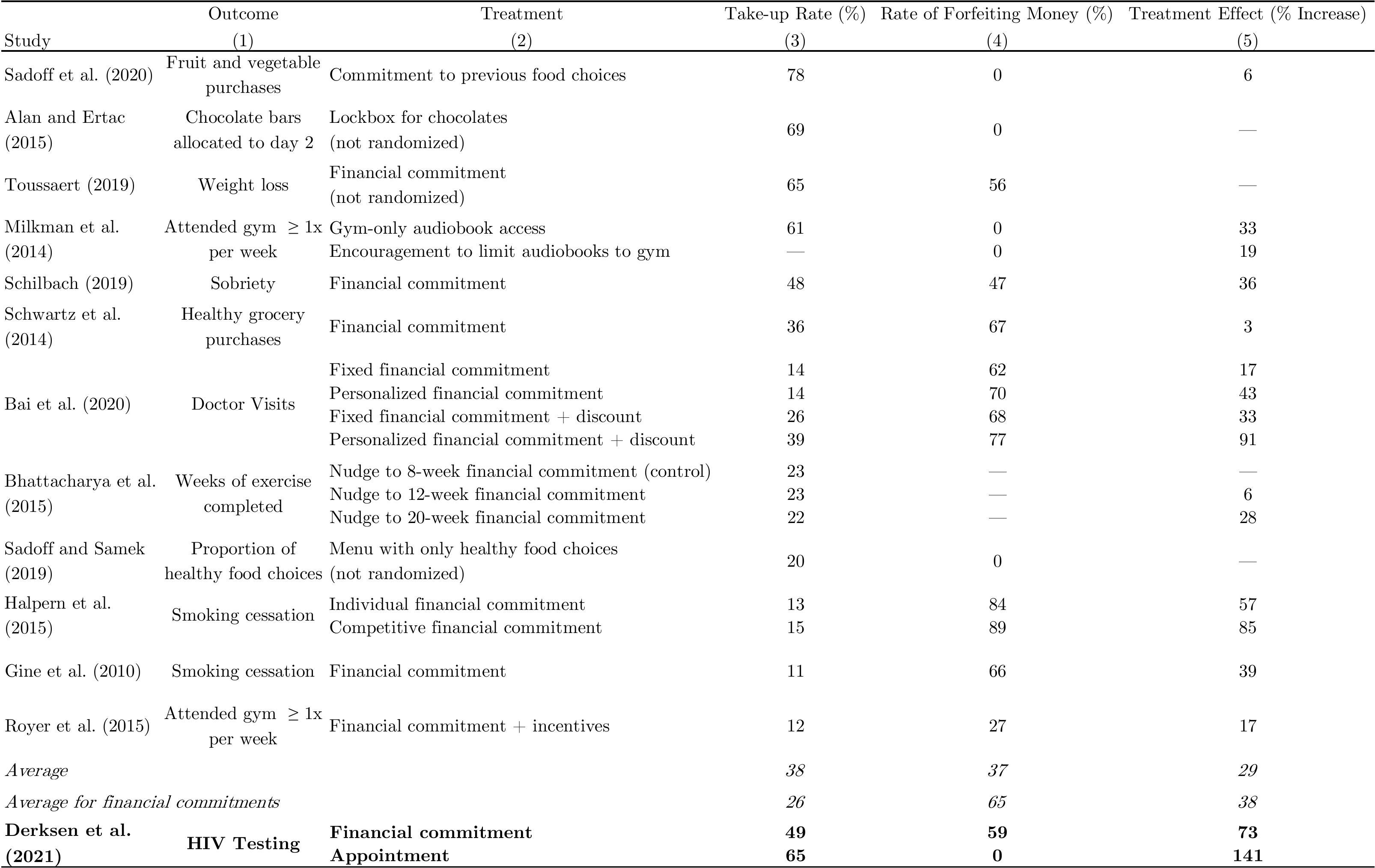}
\begin{tablenotes}[Notes]The table is sorted by the study-average value of Column 3 (the share of people offered the device who take it up). Column 4 presents the fraction of people who forfeit the money they put at stake, conditional on taking up the commitment device (if relevant). Column 5 shows the treatment effect of the commitment device on the outcome of interest, as a percent of the control-group mean. Fields marked with --- are not reported by the study. \cite{sadoff_dynamic_2020} results are for the Los Angeles site, because that is where the treatment was randomized; we show the treatment effect on fruit \& vegetable purchases, which is the only one of the seven outcomes where increases were desirable. \cite{milkman_holding_2014} take-up rate is the share of the entire sample with positive WTP in a Becker--DeGroot--Marschak mechanism. \cite{bai_self-control_2020} arm-specific forfeiture results are from personal correspondence with the authors. \cite{sadoff_can_2019} use randomized treatments to increase demand for commitment but do not look at effects on the outcome; we present the take-up rate from their control group. \cite{royerIncentivesCommitmentsHabit2015} treatment effect is the marginal effect of the financial commitment relative to the pure incentives arm.
 \end{tablenotes}
\end{center}
\vspace{-80pt}
\end{table}
\end{landscape}

This paper also contributes to the literature on appointments and behavioral nudges for health. Appointments have been shown to improve efficiency in clinics (\citealt{steenland_effects_2019}). They have also been shown to increase HIV testing rates in a lower-prevalence setting (\citealt{salvadori2020appointment}); by focusing on high-risk men in a high-prevalence region, we show that appointments are effective even when stakes are high. Appointments can be viewed as a bundle of different behavioral nudges, some of which have been studied extensively, including in the context of HIV testing.\footnote{See for example, \cite{tenthaniRetentionCareUniversal2014}, \cite{nyondoInvitationCardsPregnancy2015}, \cite{ranaShortMessageService2015}, \cite{mugoEffectTextMessage2016}, \cite{mayerMetaanalysisEffectText2017}, \cite{taylorEffectivenessTextMessaging2019}, \cite{salvadori2020appointment}, \cite{friedmanCanNudgingOvercome2021}, and \cite{macis_using_2021}.} These include studies that involve nudges to increase appointment attendance, including defaults and reminders (\citealt{chapmanDefaultClinicAppointments2016,milkman_mega-study_2021}), as well as interventions that invite participants to make a plan, or a private commitment, without offering a formal appointment (\citealt{milkmanUsingImplementationIntentions2011,kavanagh2020planning,john_can_2021}). Planning prompts appear to affect behavior in some health contexts, but in the absence of any social commitment or reminder their measured effects are often small.

Our results suggest that wider use of medical appointments in developing countries could reduce healthcare under-utilization, which is common in developing countries (\citealt{glasziou_evidence_2017}). The use of appointments in healthcare appears to be quite rare in Malawi: qualitative data from clinics in Zomba shows that only a few highly specialized services use appointments at all; those that do tend to only specify a day rather than a specific time, and do not provide reminders. Mobile phones and smartphones are now quite widespread in developing countries, making healthcare appointments technologically feasible. Appointments are a promising tool for addressing barriers to healthcare around the world, and their use should be expanded.
\section{Experiment and Data}
\label{sec_experiment}

We conducted a randomized controlled trial to assess and compare the impacts of two interventions on HIV testing: scheduled appointments and financial commitment devices. The experiment took place in the city of Zomba in southern Malawi, where the prevalence of HIV is approximately 13 percent (\citealt{national2017malawi}).  In Malawi, HIV testing, as well as most other services, are offered primarily on a walk-in basis, as opposed to by appointment.\footnote{See \ref{sec_qualappthiv} and \ref{sec_qualapptother} for further details.} HIV testing services are provided for free, and ART is provided for free to anyone who tests positive. There are 11 clinics that offer HIV testing in the city of Zomba. The study lasted for 3 months, from June 21\textsuperscript{st} to September 30\textsuperscript{th} of 2019.

Our sample consists of men from Zomba who are at high risk of HIV infection. We focus on men because they are less likely than women to be tested and treated for HIV, and more likely to die of AIDS (\citealt{dovel2015men}).\footnote{One reason women in Malawi are more likely to be in HIV care is because they receive routine tests and treatment initiation as a part of antenatal care (\citealt{nationalstatisticaloffice/malawiMalawiDemographicHealth2017}). HIV testing is supposed to be a voluntary part of antenatal care visits, but is perceived as compulsory (\citealt{angotti_offer_2011}).} Despite the fact that most Malawian men report a fairly recent clinic visit, more than half have not been tested for HIV in the past year (\citealt{dovel_missing_2020}). To construct a sample of high-risk men, we recruited participants at bars and nightclubs in Zomba. We selected these locations because they are commonly used by sex workers to find clients. We screened potential participants for mobile phone ownership, as mobile phones were essential to our implementation strategy. No potential participants were screened out for this reason; mobile phone ownership is very common in this population.\footnote{According to the 2015-16 DHS, 82.97 percent of urban men in southern Malawi owned a mobile phone; this fraction is probably even higher among men who frequent bars and engage in transactional sex.} We also screened out men who were already diagnosed and on ART treatment, as well as men who appeared to be intoxicated, who knew the interviewers personally, who did not live in the city of Zomba, or who were under 18 years old.

Men who agreed to take part in the study and satisfied the screening criteria completed a baseline survey. The survey was administered by trained enumerators, and took place near where the participant was recruited in an area that afforded privacy. Every participant was offered a MK2,000 (approximately USD \$2.76 at market exchange rates) gift of mobile phone credit at the end of the survey, and a MK500 mobile phone credit voucher that they could redeem by visiting any HIV testing clinic in Zomba within the three month study period.\footnote{To redeem the voucher, the participant had to show the testing staff a text message on their mobile phone that contained a unique identifier code. This enables us to accurately link HIV tests to study participants, and capture all HIV tests that took place in the study area. The voucher also helped to defray time or transportation costs of going to the clinic, and might have mitigated the stigma associated with HIV testing (\citealt{angelucci_adverse_2021},\citealt{derksen_love_2021}). A financial reward for getting tested gives men a different ``excuse'' to go to the clinic (\citealt{thorntonDemandImpactLearning2008}, \citealt{ngatia_social_2016}). The list of participating clinics included all 11 HIV testing clinics within and around the city of Zomba, and this list was shared with all participants. The men in the study were told that the vouchers had a two-month deadline for use, but this was not emphasized in the recruitment script. It also was not strictly enforced: HDAs were instructed to accept any vouchers presented to them while the study was ongoing, and had no way to check the date validity. We did ensure that the clinics continued to be staffed and accept men for testing for two months after the last participant was recruited into the study.} Participants did not have to agree to an HIV test to redeem the MK500 voucher; both voucher redemption and actual HIV testing were recorded.\footnote{Our main outcome variable is HIV testing, rather than simply voucher redemption; we show that our main results are robust to using any voucher redemption in Appendix \autoref{tab_voucher}.} HIV testing is free in Malawi, and was free for our study participants.

We used a factorial design that assigned men to one of four study arms: (1) a control group, (2) a group that was offered financial commitment devices only, (3) a group that was offered appointments only, and (4) a group that was offered both financial commitment devices and appointments. Randomization was done using pre-randomized lists of intervention assignments that were loaded onto the tablets used for the baseline survey.\footnote{The random assignments were linked to respondents via sequential ID numbers for each survey interviewer and day. For example, interviewer 1's first respondent on day 1 was ID \# 010101, their second respondent was 010102, etc.; each of these IDs had a pre-specified random study arm assignment linked to it.}

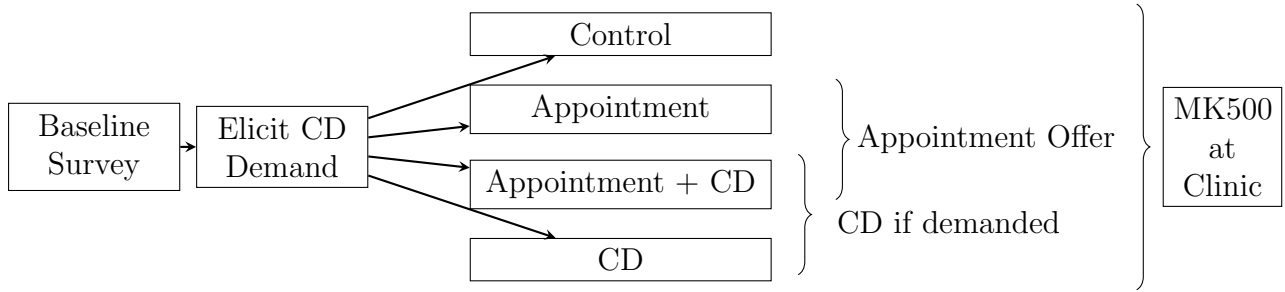
\begin{figure}[p!]
    \caption{Experimental Design \\ \footnotesize{(CD = Commitment Device)}\label{fig:experiment}}
    \centering
\begin{tikzpicture}[node distance=2cm]
\node (baseline) at (0,0) [process, text width=2cm] {Baseline Survey};
\node (elicit) at (2.5,0) [process, text width=2cm] {Elicit CD Demand};
\node (control) at (7,1.5) [startstop]{Control};
\node (appt) at (7,0.5) [startstop]{Appointment};
\node (both) at (7,-0.5) [startstop]{Appointment + CD};
\node (cd) at (7,-1.5) [startstop]{CD};
\draw[arrow] (elicit) --(control);
\draw[arrow] (elicit) --(appt);
\draw[arrow] (elicit) --(both);
\draw[arrow] (elicit) --(cd);
\draw[arrow] (baseline)--(elicit);



\draw [decorate,decoration={brace,amplitude=5pt},xshift=-4pt,yshift=0pt]
(9.5,-0.1) -- (9.5,-1.7) node [black,midway,xshift=2cm, yshift=-0.1cm] 
{CD if demanded};

\draw [decorate,decoration={brace,amplitude=5pt},xshift=-4pt,yshift=0pt]
(10,0.9) -- (10,-0.7) node [black,midway,xshift=2cm] 
{Appointment Offer};

\draw [decorate,decoration={brace,amplitude=5pt},xshift=-4pt,yshift=0pt]
(14,1.9) -- (14,-1.9) node [black,midway,xshift=2cm] 
{};
\node (voucher) at (15, 0) [process, text width=1.3cm]{MK500 at Clinic} ;

\end{tikzpicture}
  
\end{figure}

Both interventions were offered to participants at the end of the baseline survey. Before offering either intervention, we elicited demand for the financial commitment device from all participants. For men who were assigned to the combined treatment, we offered the financial commitment device first, followed by the appointment. \autoref{fig:experiment} depicts the implementation process in more detail.

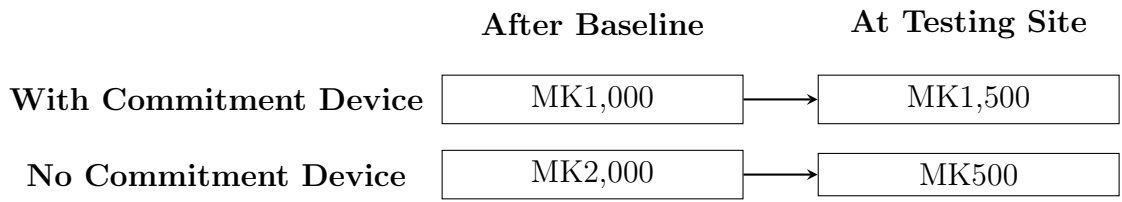
\begin{figure}[p!]
    \caption{Participant Compensation with and without Commitment Device\label{fig:voucher}}
    \centering
\begin{tikzpicture}
\node(baseline) at(0,1) [startstop2] {\textbf{After Baseline}};
\node(site) at(5,1) [startstop2] {\textbf{At Testing Site}};

\node(CD) at(-5,0) [startstop2]{\textbf{With Commitment Device}};
\node(noCD) at (-5,-1) [startstop2]{\textbf{No Commitment Device}};

\node(baselineCD) at(0,0) [startstop]{MK1,000};
\node(baselineNOCD) at(0, -1) [startstop] {MK2,000};
\node(clinicCD) at(5,0) [startstop] {MK1,500};
\node(clinicNOCD) at(5,-1) [startstop] {MK500};
 
 \draw [arrow] (baselineCD)--(clinicCD);
  \draw [arrow] (baselineNOCD)--(clinicNOCD);

\end{tikzpicture}
\end{figure}

The financial commitment device allowed participants to stake half (MK1,000) of their study compensation on visiting an HIV testing clinic. That is, rather than receiving MK2,000 immediately and MK500 upon visiting the clinic, they could choose to receive MK1,000 immediately and MK1,500 at the clinic. It was made clear that the commitment device was voluntary. Before offering the commitment device, we elicited incentivized demand for the commitment device from each participant. All participants, including those not randomized to the commitment arm, were told how the commitment device worked. Participants were then asked if they were interested in receiving such a commitment device, and were told that their choice would only be implemented with 50 percent probability, and that they would otherwise receive MK2,000 immediately regardless of their choice. After recording demand for the commitment device, we revealed the result of the random assignment. Participants who were randomized into one of the commitment device arms and who had requested the commitment device received MK1,000 at the end of the survey, as well as a MK1,500 voucher to be redeemed at any clinic. Collecting this voucher was not conditioned on agreeing to an actual HIV test; men simply had to appear at the clinic. Everyone else received MK2,000 immediately, as well as the standard MK500 clinic voucher (see \autoref{fig:voucher}). 

Participants who were randomized into one of the two appointments arms were given the opportunity to schedule an appointment at any of the 11 study clinics during regular clinic hours.\footnote{We did not mention the possibility of appointments to other participants, to avoid inducing John Henry effects (\citealt{saretsky_oeo_1972}).} Participants could choose the clinic, as well as the date and time of the appointment, starting from the day after their baseline survey and ending two and a half months after the start of the study. They could also decline the appointment. Those who chose to schedule an appointment received a phone call from testing staff two days before their scheduled appointment, reminding them of the time and place. The reminder call also allowed participants to reschedule their appointments if they wished. Note that participants in the appointments arms still received MK500 clinic vouchers which could be redeemed at any clinic and at any time; they did not need to attend their appointment to redeem this voucher, and there was no financial penalty for missing the appointment.

\subsection{Data}
\label{sec_data}

Our study uses four sources of data: a baseline survey, administrative data on HIV testing collected in all clinics offering HIV testing within and around the city of Zomba, records of appointment reminder calls, and qualitative interviews with clinic staff.

After recruiting participants, our team of interviewers conducted a baseline survey, which included questions on demographic characteristics (gender, ethnicity, age, place of birth, educational achievement, marital status), socio-economic characteristics (employment, income, assets, expenditures), sexual behaviors (number of partners, risk, perceptions), past HIV tests, and intentions to get tested for HIV in the future. The baseline data was used for three purposes: (1) as a source of control variables to increase statistical power, (2) to explore mechanisms via heterogeneous treatment effect analyses, and (3) to prevent impostors from claiming the vouchers on behalf of our participants, and to remove those who did from the data. 

To collect HIV testing data, we collaborated with the 11 clinics that provide HIV testing services within and around the city of Zomba. We hired qualified HIV Diagnostic Assistants (HDAs) and integrated them into each clinic.\footnote{In some cases we placed new HDAs in the clinic, and in others we directly hired existing clinic staff to work for our team for the duration of the project.} This was done to ensure that our participants would not face wait times for their appointments, as well as to avoid disruptions to the clinic's usual operations due to our study. In any case, demand for HIV testing is low at all clinics in Zomba, and at the time of the study HIV test seekers did not typically have to wait.\footnote{See \ref{sec_qualwait} for information on wait times from qualitative interviews with clinic staff.}

To ensure participants' privacy, and to make them more comfortable with revealing personal information about their sexual behavior and HIV testing decisions, we did not collect any names as part of the study. Instead, we verified participants' identities at the time of voucher redemption by requiring them to show testing staff the text messages they received on their mobile phones during the baseline survey, which included unique voucher codes. These voucher codes were tied to the answers to specific security questions we asked during the baseline survey.\footnote{We used three questions: the name of the participant's primary school, their year of birth, and their mother's district of birth. To reduce the likelihood of intentional impersonation, these security questions were inconspicuous to participants, and we did not specify how participants' identities would be verified at the clinic. The HDAs verified the answers to these security questions at the time of voucher redemption. Just ten of the participants who got an HIV test gave more than one incorrect answer to our security questions, which was our pre-specified criterion for labeling a test-taker as an impostor. We code the outcome variables for these impostors as zeroes rather than ones; our results are also robust to coding their outcomes as ones (see Appendix \autoref{tab_impostors}).}

Our project's HDAs were in charge of collecting vouchers, recording data, and performing HIV tests for study participants. In line with local protocols, those who tested negative were encouraged to seek a second confirmatory test after 3 months, and the HDAs worked with the clinics to integrate newly-diagnosed individuals into ART initiation and care. The HDAs recorded data in handwritten notebooks. This included participants' voucher codes and phone numbers, in order to link them to our baseline data, as well as whether they agreed to be tested for HIV. HIV test results and ART initiation were  recorded only by study arm, and are not linkable to individual participants. The information in the notebooks was digitized only at the end of the experiment.

Our sample consists of 1,232 men.\footnote{Consistent with our pre-analysis plan, our analysis excludes participants who named a clinic outside of the study area as their preferred clinic to do an HIV test, to ensure that we observe all HIV tests that take place post-intervention.} Of those, 301 participants were assigned to the control group, 329 were offered only appointments, 295 were offered only commitment devices, and 307 were offered both appointments and commitment devices. Descriptive statistics and balance tests are provided in Appendix \autoref{tab_balance}. The average age of the men in our sample is just under 32 years old. Roughly 98 percent of our sample can read and write, and the average participant had finished just under 11 years of schooling. In our sample, 66 percent of participants are married and 43 percent report having a girlfriend.  The most common ethnic group is Lomwe (30 percent) followed by Yao (21 percent) and then Chewa and Ngoni (13 percent each). Over 90 percent of participants report having been tested for HIV at least once, and the average man in the sample reports having been tested for HIV five times. 

The control and treatment groups are well-balanced on baseline variables. Omnibus $F$-tests of the joint significance from regressions of the treatment indicators on all the baseline covariates yield $p$-values that are above usual significance thresholds. As an alternative to significance tests for balance, the table also shows pairwise normalized differences across study arms (\citealt[p. 310]{imbens_causal_2015}). These are small: out of 68 differences, 66 are below 0.1 and all are below 0.15, which indicates good balance. As our outcome variables are constructed using administrative data, our study does not suffer from problems of attrition due to respondents not being found or not consenting to participate in a second survey. 

To better understand the study context, especially the use of health appointments and healthcare wait times in Malawi, we also collected qualitative data from clinics in the country. We interviewed staff at all clinics in the study area, as well as four other large clinics in Zomba District, and seven hospitals from different regions of Malawi. We interviewed both clinic coordinators and HDAs.\footnote{The qualitative data collection was approved separately from the main study, by IRBs at the Malawi College of Medicine Research Ethics Committee (2572/2019) and at the University of Toronto (36913/2020).}

We also conducted a follow-up survey in 2020, to study longer-run treatment effects and analyze beliefs and behaviors related to COVID-19 (\citealt{fitzpatrick_health_2021}). For this survey, we attempted to contact all 1,232 men from our original sample by phone. However, our effective attrition rate (after removing impostors) was 81 percent because people in Malawi frequently change SIM cards. We are therefore unable to draw meaningful conclusions from this data.
\section{Empirical Strategy}
\label{sec_empirical_strategy}

Our empirical strategy is based on a pre-analysis plan that we filed prior to receiving the outcome data (\citealt{derksen_soft_2019}).\footnote{The analysis plan can be accessed at \url{https://www.socialscienceregistry.org/versions/57507/docs/version/document}.} We discuss several minor deviations from the pre-analysis plan below in \autoref{sec_deviations}; these deviations increase the rigor of the analysis and do not alter our substantive conclusions. 

Our primary outcome of interest is the decision to get an HIV test. This is captured by an indicator $T_i$ that is equal to one for men who redeemed their voucher at one of the study clinics and followed through with an HIV test, and zero otherwise.

\subsection{Intention-to-Treat Estimates}
\label{sec_itt}

To estimate the intention-to-treat effects of the two treatments on HIV testing, we use linear regressions of the following form:
\begin{align}
T_i = \alpha + \beta_1 A_i + \beta_2 CD_i + \beta_3 A_i \times CD_i + X_i^\prime\gamma +\varepsilon_i 
\label{eq_itt}
\end{align}
\noindent where $A_i$ indicates a participant was assigned to the appointments treatment and $CD_i$ indicates assignment to the commitment device treatment. $A_i \times CD_i$ is the interaction between the two treatments.\footnote{In Appendix \autoref{tab_itt_short} and \ref{sec_short_specification}, we also show the ``short'' specification that omits this interaction term. The treatment effect estimates using the  short specification are biased downward because this specification ignores the important negative interaction between the two treatments (\citealt{muralidharan_factorial_2019}).} $X_i$ is a list of baseline characteristics, which are included in the regression to increase precision.

We consider three different sets of control variables $X_i$. Our first specification includes no controls at all. In the second specification, we control for the variables and fixed effects that we specified in our pre-analysis plan. We include 10 variables that were significant predictors of past testing behavior as recorded in the baseline survey (these variables are described in \autoref{tab_controls_prespec}). We also include fixed effects for the date of the baseline survey, the baseline survey interviewer, and the participant's preferred testing clinic as reported at baseline. We focus on the results from this second specification, and show that our results are robust to varying the controls we use. The third specification includes all the pre-specified fixed effects, but selects other controls using the double LASSO method of \cite{chernozhukov2017double}. Specifically, we use the \texttt{pdslasso} command in Stata (\citealt{ahrens_pdslasso_2019}), and ask it to select variables from \autoref{tab_controls_prespec} and \autoref{tab_controls_other}.\footnote{In addition to linear terms for each variable, we follow \cite{knaus_heterogeneous_2020} in feeding the algorithm fourth-degree polynomials and logarithmic terms in each variable, along with all first-order interactions between variables. For variables with zeros or negative values, instead of using the logarithm, we use both the inverse hyperbolic sine transformation (\citealt{burbidge_alternative_1988}) and the \cite{ravallion_concave_2017} concave log-like transformation. Prior to variable selection, we partial out fixed effects for the date of the baseline survey, the baseline survey interviewer, and the participant's preferred testing clinic as reported at baseline. The exact variables chosen by the double LASSO vary by specification but, for our main treatment effects analysis (\autoref{tab_itt}, Column 3), it selects three variables: 1) the Ravallion transformation of the index of willingness to get an HIV test, 2) the interaction between ever having been tested and demand for the commitment device, and 3) the interaction between the number of sexual partners in the past 12 months and the demand for the commitment device.}

The random assignment of participants to study arms ensures that $\hat{\beta_1}$ and $\hat{\beta_2}$ are consistent estimates of the intention-to-treat effects of the two treatments (when offered on their own), provided that the stable unit treatment value assumption holds---that is, that there are no spillovers between the men in the study. This is a plausible assumption in our case: the men were interviewed individually and privately, and there was no direct way for them to share the interventions with one another nor any incentive for them to do so. 


\subsection{Treatment-on-the-Treated Effects}
\label{sec_TOT}

Take-up of the two treatments was imperfect: not everyone offered an appointment took one, and not everyone assigned to the commitment device arm wanted one. We thus estimate the treatment-on-the-treated effect of each intervention---the average effect of the intervention on the people who actually used it. For the appointments intervention, we do this by defining an indicator $A^{TOT}_i$, which is equal to one if the participant signed up for an appointment and zero otherwise. We then estimate the following equation via 2SLS, using $A_i$ as an instrument for $A^{TOT}_i$, and restricting the sample to the control group and the appointments-only arm:\footnote{We restrict the sample to focus on the treatment-on-the-treated effect of the appointments-only treatment (i.e., without needing to consider the interaction with the commitment device treatment).}

\begin{align}
T_i = \alpha + \beta_1 A^{TOT}_i + X_i^\prime\gamma +\varepsilon_i
\label{eq_2sls}
\end{align}

For the commitment devices, we know who the compliers are because we elicited demand for commitment devices for all the men in the study. We can therefore define indicators $D_i$ for men who demand the commitment device and $(1-D_i)$ for men who do not, and estimate the treatment-on-the-treated effect of the commitment devices using the following equation:

\begin{align} \label{eq_hte_o}
T_i = &D_i \times (\beta_0 + \beta_1 A_i + \beta_2 CD_i + \beta_3 A_i \times CD_i) \nonumber\\
&+ (1-D_i) \times (\beta_4  +\beta_5 A_i ) +  X_i^\prime\gamma +\varepsilon_i 
\end{align}

The first part of Equation \eqref{eq_hte_o} examines the effects of the two treatments on men who demand the commitment device. For these men, $\beta_0$ measures the average testing rate in the control group, $\beta_1$ is the intention-to-treat effect of the appointments-only treatment, $\beta_2$ is the treatment-on-the-treated effect of the commitment device treatment (i.e., the effect of the commitment device on men demanding it), and $\beta_3$ captures the interaction effect of the two treatments. 

The second part of Equation \eqref{eq_hte_o} focuses on men who do not demand the commitment device. For these men, the variable $CD_i$ is irrelevant as none of them received the commitment device (and they were not even told whether they would have been assigned to the commitment device arm or not). For this group of men, the control-group testing rate is given by $\beta_4$ and the intention-to-treat effect of the appointments treatment is given by $\beta_5$. 


\subsection{Exploratory Analyses}
\label{sec_exploratory}

We conduct three exploratory analyses on top of our main analysis. First, we conduct conventional analyses of treatment effect heterogeneity by interacting the two treatment indicators with baseline covariates. We estimate the following specification:

\begin{align}
T_i = \alpha &+ \beta_1 A_i + \beta_2 CD_i + \beta_3 A \times CD_i \nonumber\\
& + \beta_4 A_i \times W_i + \beta_5 CD_i \times W_i \\
\nonumber &+ \beta_6 A_i \times CD_i \times W_i + \beta_7 W_i + X_i^\prime\gamma +\varepsilon_i 
\label{eq_heterogeneity}
\end{align}

\noindent where $W_i$ is the baseline covariate of interest. In addition to estimating these interactions separately, we also conduct a pooled analysis in which we include all of the variables simultaneously, while also controlling for main effects and interactions with the treatments for every variable in \autoref{tab_controls_prespec}. We de-mean all the variables $W_i$ prior to constructing the interaction terms, so the main effects of the treatments can still be interpreted as average treatment effects (\citealt[p.~247]{imbens_causal_2015}).

Second, we investigate heterogeneous treatment effects by propensity to seek an HIV test, using the repeat split-sample (RSS) endogenous stratification procedure of \cite{abadie2018endogenous}. The approach relies on randomly splitting the sample in half, and using half of the data to predict the outcome variable (the first stage) and the other half to use the predicted outcomes for treatment effect heterogeneity analysis (the second stage).\footnote{We use the  \texttt{estrat} Stata command (\citealt{ferwerda_estrat_2014}). We conduct 100 random splits of our sample and do the two stages for each; our point estimates and standard errors come from the mean and standard deviation of the estimates from stage 2 across the 100 sample splits. In the first stage of the procedure, we use the two different sets of control variables from \autoref{eq_itt} as predictors, not including the fixed effects. For the double LASSO controls, to avoid different controls being selected for each sample split, we use the predictors selected for estimating \autoref{eq_itt}.
In the second stage, we estimate treatment effects for terciles of the predicted outcome calculated in the first stage, using the fixed effects as controls. We run the procedure just on the control group and the appointments-only arm for the analysis of the appointments treatment, and just on the control group and the commitment devices-only arm for the analysis of the commitment device treatment.}

Finally, we study the impacts of the treatments on indicators for testing positive for HIV and for initiating antiretroviral treatment, since getting HIV-positive people into treatment is a major goal of current testing campaigns. These outcomes are anonymized and linked only to the study arm the participant was in, so we cannot include covariates in regressions with these outcomes. Note that our study was not powered to study these outcomes, because being diagnosed with HIV is a rare outcome (in our study's control group, just 11 percent of men got a test and only 6 percent of HIV tests were positive, so less than 1 percent of the control group tested positive for HIV).

\subsection{Inference}
\label{sec_inference}

Our inference is based on conventional Eicker--Huber--White (EHW)  heteroskedasticity-robust standard errors, with no adjustment for clustering, because our treatment was randomized at the individual level (\citealt{abadie_when_2017}). The uncertainty in our estimates comes from the randomization of the treatment, rather than sampling variation. However, the conventional sampling-based standard errors that we report will be conservative on average, i.e., larger than the correct standard errors that capture design-based uncertainty (\citealt{abadie_sampling-based_2020}). We also show that our main results are robust to randomization inference. 

To address multiple testing concerns we compute sharpened $q$-values that control the false discovery rate (FDR) using the \cite{benjamini2006adaptive} two-step procedure. We use \cite{anderson2008multiple}'s approach and report the lowest value for which the \citeauthor{benjamini2006adaptive} approach rejects the null hypothesis, so our $q$-values are comparable to $p$-values.

We conduct our FDR corrections across all the $p$-values for treatment effects reported in the paper and appendix. This includes Tables \ref{tab_itt} through \ref{tab_hte_by_commitment} and Appendix Tables \ref{tab_itt_short} through \ref{tab_impostors}, as well as the tests associated with Figures \ref{figure_delays_before_testing} and \ref{figure_delays_by_appt_date_vs_no}. It does not include any of the hypothesis tests in the balance table (Appendix \autoref{tab_balance}). The tables show significance stars based on $p$-values, and sharpened $q$-values in brackets. All discussions of statistical significance in the text are based on the $q$-values.

\subsection{Deviations from the Pre-Analysis Plan}
\label{sec_deviations}

Our analyses deviate from the pre-analysis plan in several ways. First, while we did pre-specify a plan to examine the effect of the interaction term $A_i \times CD_i$ as a secondary analysis, our pre-specified analyses mainly did not include this interaction term. This ``short'' model has higher power when the interaction term does not matter, but leads to incorrect estimates when it does (\citealt{muralidharan_factorial_2019}). Since our results show that the interaction between the two treatments is important, we focus on the fully-interacted specifications. However, none of our inferences or qualitative conclusions depend on this choice. We present the complete tables using the exact specifications from the pre-analysis plan in \ref{sec_short_specification}.

Second, we handle the control variables slightly differently than specified in the pre-analysis plan. Our preferred specification, with our pre-specified list of controls, matches the analysis plan exactly. For the double LASSO, the analysis plan called for the fixed effects to be included in the selection procedure; we instead partial them out in advance. The analysis plan also did not specify the construction of the higher-order and logged terms and interactions. In addition to the pre-specified and double LASSO controls, we also show a specification with no controls, which was not in the analysis plan.  

Third, the analysis plan specified that we would conduct the FDR adjustments only when testing related hypotheses, and not across estimands, and that the exploratory analyses would not be included in the adjustments at all. We take a broader approach, conducting the FDR procedure across all $p$-values included in the paper (including in the appendices).

Fourth, we conduct additional exploratory analyses that were not listed in the analysis plan. In particular, our analysis of the timing of test dates in Section \ref{sec_mechanisms} was not part of our analysis plan; rather, it is an exploration of a potential mechanism that we first conceived of after we had already seen the main results.

Note that the follow-up survey mentioned at the end of \autoref{sec_data} was not part of our pre-analysis plan. We also did not file a separate analysis plan for this data, intending to treat any analyses of the data as exploratory. As noted in that section, we do not use any of that data in this paper.
\section{Results}
\label{sec_results}

Demand for both appointments and commitment devices is quite high in our setting. Among the men randomly offered an appointment, 65 percent signed up for one. We elicited demand for commitment devices for the entire sample: 51 percent of all the men in the study wanted a commitment device, including 49 percent of men assigned to one of the commitment device study arms. The high take-up of the commitment device suggests that a majority of the people in the study were aware of their self-control issues when it comes to HIV testing, and believed the intervention could help them overcome these issues. However, for many study participants this sophistication about self-control problems was only partial: just 41 percent of men who signed up for a commitment device actually followed through and visited a clinic. As a result, 59 percent of men who enrolled in the commitment device simply lost their investment. This parallels the results in \cite{bai_self-control_2020} and \cite{john_when_2020}, who also find substantial failures to follow through on commitment devices---and thus find that people are made worse off by the offer of a commitment device.

\begin{figure}[p!]
\centering
\caption{HIV Testing Rates by Study Arm\label{fig_testing_by_arm}}
\includegraphics[width=\textwidth,page=1]{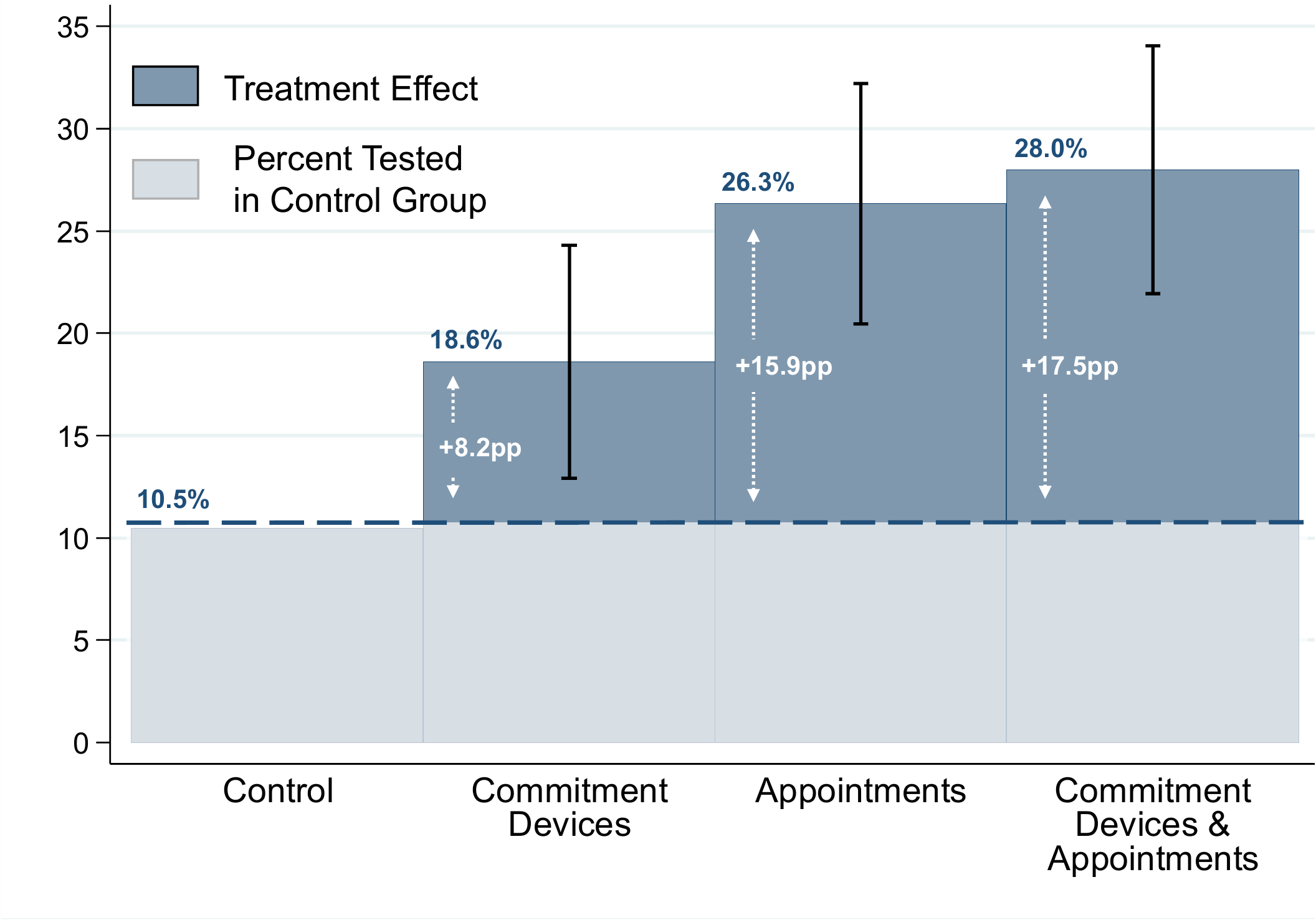}
\begin{figurenotes}[Notes]Bars represent predicted values for each study arm based on Column 2 of \autoref{tab_itt}, which uses \autoref{eq_itt} with our pre-specified control variables and the full set of fixed effects. 
\end{figurenotes}
\end{figure}

Both interventions significantly increase HIV testing. \autoref{tab_itt} shows the intention-to-treat effects of the two interventions on HIV testing. The bar chart in \autoref{fig_testing_by_arm} illustrates the results, showing testing rates by study arm based on our preferred specification (Column 2 of \autoref{tab_itt}). Appointments, offered on their own, cause a 16 percentage-point increase in HIV testing---a 141 percent increase relative to the control group mean. This effect is statistically significant at the 0.01 level. Appointments are much more effective than commitment devices at increasing HIV testing. Commitment devices, offered on their own, increase HIV testing by eight percentage points---about half of the effect of the appointments treatment. This effect is also statistically significant at the 0.01 level. We can reject the equality of the appointment and commitment device effects at the 0.05 level.

\begin{table}[p!]
\caption{Effects of Appointments and Commitment Devices on HIV Testing\label{tab_itt}} 
\includegraphics[width=\linewidth,page=2]{Tables/Appointments_Tables_cropped.pdf}
\begin{tablenotes}[Notes]Sample is 1,232 men who completed a baseline survey. Pre-specified controls include all the variables in \autoref{tab_controls_prespec}. Double Lasso controls uses the \cite{chernozhukov2017double} method for variable selection and inference, as described in \autoref{sec_itt}. Columns 2 and 3 also control for date-of-survey fixed effects, enumerator fixed effects, and preferred clinic fixed effects. Heteroskedasticity-robust standard errors in parentheses: * $p<0.1$; ** $p<0.05$; *** $p<0.01$; \cite{anderson2008multiple} sharpened $q$-values in brackets.
\end{tablenotes}
\vspace{-10pt}
\end{table}

The coefficient on the interaction term between the two treatments is large and negative, showing that appointments and commitment devices are substitutes. Its $q$-value is slightly above the conventional significance threshold of 0.1 in our preferred specification (Column 2). The interaction term is almost as large in magnitude as the main effect for the commitment device treatment, and we cannot reject that the sum of the two coefficients is equal to zero ($q$-value = 0.35). In other words, getting the commitment device on top of an appointment generates next to zero additional effect on testing rates. In contrast, getting an appointment on top of the commitment device has a marginal effect of 9 percentage points, which is statistically significant at the 0.05 level. While appointments are strong substitutes for commitment devices, commitment devices are imperfect substitutes for appointments.

We estimate the treatment-on-the-treated effects of appointments in \autoref{tab_2sls}. The average treatment effect of an appointment on men who actually signed up for one is a 24 percentage-point increase in HIV testing. 

\begin{figure}[p!]
\caption{Variation in Treatment Effects by Demand for,\\and Random Assignment to, the Commitment Device\label{fig_HTE_by_CD_demand}}
\includegraphics[width=\textwidth,page=2]{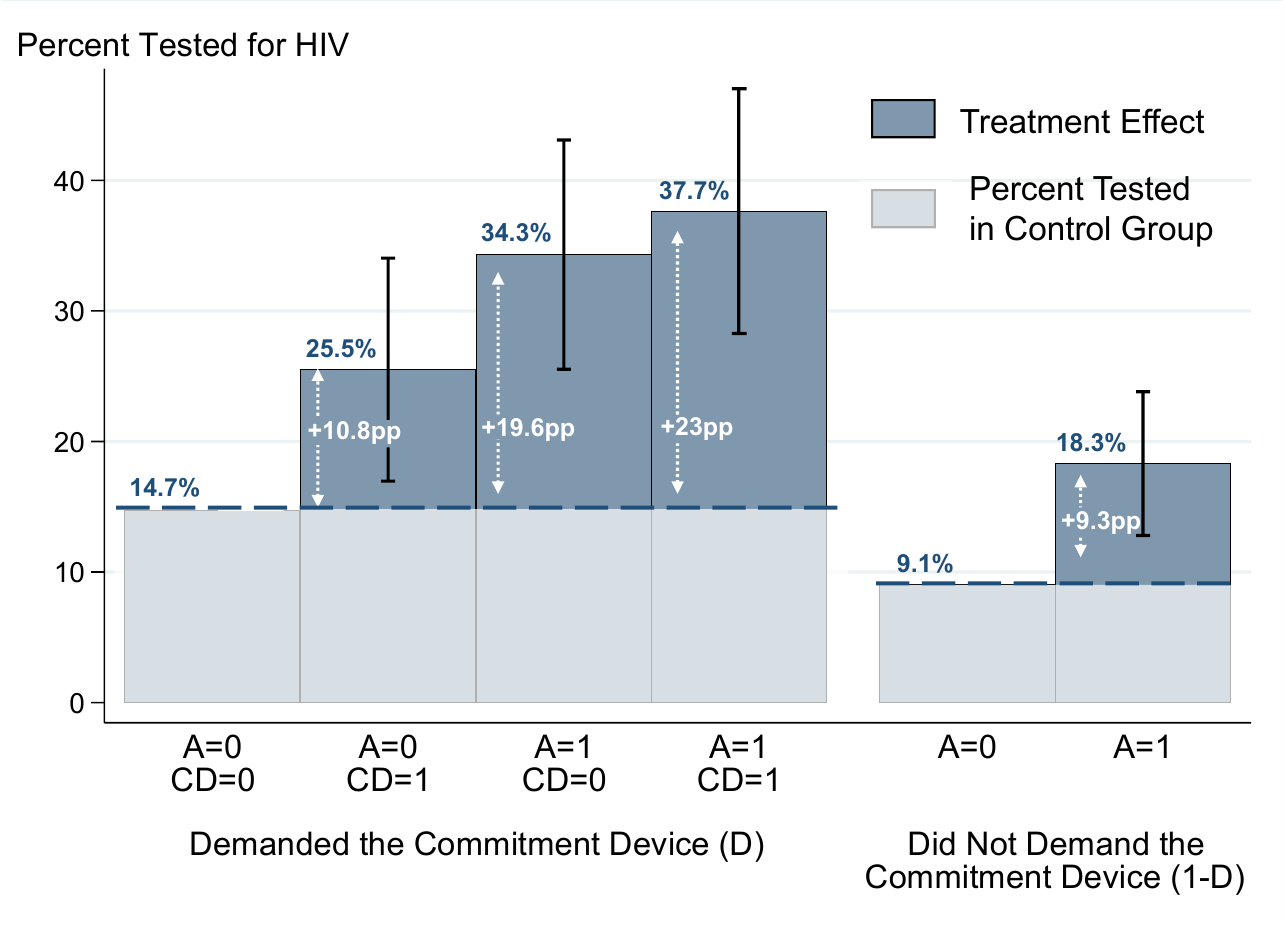}
\begin{figurenotes}[Notes]Predicted values of the outcome variable based on Column 2 of \autoref{tab_hte_by_commitment}. $CD=1$ for participants who were randomly assigned to one of the commitment device arms and $CD=0$ otherwise. $A=1$ for participants who were randomly assigned to one of the appointment arms and $A=0$ otherwise. Whiskers show 95 percent confidence intervals for the differences between the treatment and control bars. Arrows show the magnitudes of the treatment effects.
\end{figurenotes}
\end{figure}

We estimate the treatment-on-the-treated effect of the commitment device treatment in \autoref{tab_hte_by_commitment}.\footnote{\autoref{tab_hte_by_commitment} focuses on the long specification. Results for the equivalent short specification are shown in Appendix \autoref{tab_hte_by_commitment_short}.} The results are also shown in \autoref{fig_HTE_by_CD_demand}. Recall from equation \eqref{eq_hte_o} that this specification does not include a constant, so the first and the fifth rows of \autoref{tab_hte_by_commitment} represent absolute testing levels in the control arm. The effect of a commitment device on men who demand one is between 11 and 13 percentage points.

\begin{table}[htbp!]
\centering
\caption{2SLS Estimates of the Treatment-on-the-Treated Effect of Appointments\label{tab_2sls}} 
\includegraphics[page=3]{Tables/Appointments_Tables_cropped.pdf}
\begin{tablenotes}[Notes]Sample is 1,232 men who completed a baseline survey. Pre-specified controls include all the variables in \autoref{tab_controls_prespec}. Double Lasso controls uses the \cite{chernozhukov2017double} method for variable selection and inference, as described in \autoref{sec_itt}. Columns 2 and 3 both control for date-of-survey fixed effects, enumerator fixed effects, and preferred clinic fixed effects. Panel B shows the effective $F$-statistic of \cite{montiel_olea_robust_2013}. Heteroskedasticity-robust standard errors in parentheses: * $p<0.1$; ** $p<0.05$; *** $p<0.01$; \cite{anderson2008multiple} sharpened $q$-values in brackets.
\end{tablenotes}
\end{table}

\begin{table}[htbp!]
\centering
\caption{Treatment Effect Heterogeneity by Demand for Commitment\label{tab_hte_by_commitment}} 
\includegraphics[width=0.85\linewidth,page=4]{Tables/Appointments_Tables_cropped.pdf}
\begin{tablenotes}[Notes]Sample is 1,232 men who completed a baseline survey. These regressions do not include a constant; (CD) x (1-D) is omitted and assumed to be zero. Pre-specified controls include all the variables in \autoref{tab_controls_prespec}. Double Lasso controls uses the \cite{chernozhukov2017double} method for variable selection and inference, as described in \autoref{sec_itt}. Columns 2 and 3 also control for date-of-survey fixed effects, enumerator fixed effects, and preferred clinic fixed effects. All controls are standardized prior to running the regressions. Heteroskedasticity-robust standard errors in parentheses: * $p<0.1$; ** $p<0.05$; *** $p<0.01$; \cite{anderson2008multiple} sharpened $q$-values in brackets.
\end{tablenotes}
\vspace{-80pt}
\end{table} 

\autoref{tab_hte_by_commitment} also allows us to examine heterogeneity in the intention-to-treat effect of the appointments treatment by demand for the commitment device. The table shows that the appointments treatment increases HIV testing by nearly 20 percentage points among the men who wanted a commitment device but did \textit{not} receive one. This is more than twice the effect of the appointments treatment on men who did not want a commitment device---even though both groups received the exact same intervention (appointments alone, without commitment devices). We can reject the equality of the two effects at the 0.1 level. We conclude that appointments are more effective for men who have self-control issues that they are (at least partially) aware of. Moreover, for the men who did want a commitment device, appointments are nearly twice as effective as financial commitment devices.

The results of \autoref{tab_hte_by_commitment} also confirm that the two treatments are substitutes. For men who wanted a commitment device, getting both a commitment device and an appointment yields about the same total effect on HIV testing as getting only an appointment, and we cannot reject that the two effects are equal ($q$-value $\approx0.3$).

Appointments are highly cost effective. At a cost of \$0.43 per person they increase testing by 16 percentage points (\autoref{tab_cost_effectiveness}). The cost per additional person tested is \$2.69, as compared with \$3.01 for the commitment devices.\footnote{For details of the cost-effectiveness calculations, see \ref{sec_costs}. Our primary cost-effectiveness calculations show only the incremental cost of each intervention. Crucially, this omits the cost of the MK1,000 of the survey gift that was used as the commitment device, because this gift was provided to all participants. If this amount is included in the cost of the commitment device then the commitment devices cost \$19.61 per additional person tested (\autoref{tab_cost_effectiveness}, Column 3). This larger number may better reflect the true cost of implementing a policy where participants are offered gifts to use as commitment devices (as opposed to using their money). Our cost calculations also exclude the cost of finding and contacting men for the survey, since that was done in the control group as well.} A simple basis for comparison with these results is directly paying people to get tested. \cite{thorntonDemandImpactLearning2008} did something quite similar to this, testing her entire sample and paying people to pick up their results at a clinic. Cash incentives increase the rate of picking up one's results by 9.1 percent per dollar of incentive, so each additional person tested cost \$10.99.\footnote{Another point of comparison is \cite{macis_using_2021}, who find that cash incentives greatly increase HIV testing---but have much smaller effects on actually learning one's results. Our results are more comparable to Thornton's because the test results were revealed immediately.} Our increase is also measured off of a lower base rate of HIV testing (11 percent in our study vs. 34 percent in \citeauthor{thorntonDemandImpactLearning2008}'s study). In percentage terms, appointments increase testing by 141 percent; to achieve the same relative increase by paying people to learn their HIV status would cost \$5.27 per person, as compared with \$0.43 for appointments.

\begin{table}[htbp!]
  \caption{Cost Effectiveness\label{tab_cost_effectiveness}} 
 \includegraphics[width=\linewidth,page=5]{Tables/Appointments_Tables_cropped.pdf}
\begin{tablenotes}[Notes]Sample is 1,232 men who completed a baseline survey. Costs in this table are the marginal cost of running each intervention relative to the control group, and thus do not include the costs of the HIV testing vouchers. We compute costs based on the additional time needed to recruit participants in each of the two pure treatment arms, priced at the wage we paid to the survey enumerators. For the appointments, we also include the time and airtime costs of the appointment reminders, and the rescheduling calls (for the respondents who needed to reschedule). Column 3 includes the MK1,000 that participants were allowed to stake on the commitment device, since the commitment devices cannot be offered without incurring this cost. The costs of the tests themselves are not included, as the goal is to encourage the uptake of HIV tests. The effect sizes come from the main results, in Column 2 of \autoref{tab_itt}.
    \end{tablenotes}
\end{table}

\subsection{Exploratory Analyses}

We explore variation in the treatment effects of the two interventions in two different ways. First, we present conventional heterogeneous treatment effect analyses in Appendix \autoref{tab_hte}. We see no evidence of statistically-significant heterogeneity in treatment effects by men's perceived probability of having HIV, their self-reported tendency to postpone HIV testing, the extent to which they say they live for today, or their perceived benefits from taking ART. 

Second, in Appendix \autoref{tab_estrat}, we use the \cite{abadie2018endogenous} method to study how the treatment effects vary by propensity to get an HIV test.\footnote{The \texttt{estrat} command returns individual standard errors for each tercile, but not for the differences between them. We thus compute the standard errors of the differences as $\sqrt{Var(X)+Var(Y)−2Cov(X,Y)}$. We treat the covariance term as zero, as it is also not reported by the command, and thus our $p$-values are likely to be conservative.} Both appointments and commitment devices work better for men who are more likely to get tested for HIV in the absence of either intervention. In many settings, interventions are viewed as more impactful if they target demographics for whom outcomes are otherwise low. But HIV testing is a somewhat unusual outcome in this regard; those who are unlikely to test without the intervention may be at lower risk infection, and have little to gain from an HIV test. Data from Zomba suggests this is the case in our setting: HIV test-positivity rates are much higher than the local prevalence of the virus (\citealt{derksen_love_2021}), so people who are likely to get tested for HIV under the status quo are at higher risk of HIV infection. This implies that the intervention may be successfully targeting high-risk men.

Finally, we explore treatment effects on positive HIV tests and on ART initiation. Because both interventions have large effects on HIV testing, they may also increase the detection of actual HIV cases and guide HIV-positive men into treatment for the disease. We are underpowered to study these very rare outcomes, but out of the 13 men who tested positive in our experiment, 10 had been offered an HIV testing appointment. We examine treatment effects on positive HIV tests and ART initiation in Appendix \autoref{tab_hivpos_art}. The commitment devices had no significant effect on either outcome, but the appointments increase both the rate of HIV detection and the rate of ART initiation by 1 percentage point---more than doubling the control-group rates. Using the short specification, these effects are statistically significant at the 0.1 level and the effect on positive tests is essentially robust to FDR correction (the $q$-value is 0.104 for ART initiation). However, these results are not significant in the long specification, which has lower statistical power (\citealt{muralidharan_factorial_2019}). We therefore interpret this as suggestive---but not conclusive---evidence that appointments not only increase HIV testing rates, but also help to locate HIV-positive men and bring them into the treatment pipeline.

\subsection{Robustness checks}

Our primary robustness check is presented in each of the tables in the paper: all our results are basically unchanged by our choice of controls, or by using no controls at all. Columns 1 and 3 of \autoref{tab_itt} through \autoref{tab_hte_by_commitment} show results that are nearly identical to our preferred specification in Column 2, with occasional minor changes in the coefficients and $q$-values. Our main results from \autoref{tab_itt} are also qualitatively similar when we use the ``short'' specification that omits the interaction term (Appendix \autoref{tab_itt_short}), but smaller in magnitude because omitting the interaction term biases the coefficients downward (\citealt{muralidharan_factorial_2019}).

We also show our findings are qualitatively robust to switching all our analyses  to the exact approach we pre-specified in our analysis plan. \ref{sec_short_specification} shows all the results of the original specifications from our analysis plan, which are primarily focused on the short regression that omits the interaction between the two treatments. All our findings are qualitatively identical, and our inferences are substantively unchanged.

Our inferences are robust to the use of randomization inference instead of $p$-values based on Eicker-Huber-White standard errors. Appendix \autoref{tab_ri} shows randomization inference $p$-values for our main results from Columns 1 to 3 of \autoref{tab_itt}. Some of the $p$-values are somewhat larger, and others are slightly smaller, but we continue to reject a zero treatment effect for each intervention separately, and the equality of the two interventions, at the 0.05 level.

We also show that our results are robust to different ways of defining our outcome variable. One alternative definition is to use any voucher redemption, without conditioning on an HIV test. A meaningful proportion of participants visited a clinic to redeem their voucher without getting an HIV test: about 5 percent of the control group, 6 percent of men in the appointments-only arm, 9 percent of men in the commitment devices-only arm, and 9 percent of men in the combined arm. Columns 1 to 3 of \autoref{tab_voucher} show treatment effects on any voucher redemption. The estimated effect of the appointments treatment is nearly unchanged, while the effect of the commitment device treatment is somewhat larger; we can no longer reject the equality of the two effects in this table. Columns 4 to 6 show that appointments had no effect on voucher redemption alone, with no HIV test, while commitment devices increased it by four percentage points.

It appears that some men, after requesting the commitment device, come to the clinic to redeem the voucher but do not get an HIV test. This suggests another margin of partial sophistication by the men who enrolled in the commitment device. While this subset of men (who comprise eight percent of all the people who enrolled in a commitment device) did not lose their investments, they wasted their time and effort in coming to the clinic but not actually following through with a test, likely due to fear of learning their HIV status.\footnote{It is possible that some of these men did not want an HIV test, but did want to collect the MK500 from their voucher for appearing at the clinic, and used the commitment device to encourage themselves to do so. Another possible non-testing motivation for using the commitment device is as a savings instrument. While the interest rate on these savings is zero, \cite{brune_pay_2021} find high demand for a zero-interest deferred-payment savings product elsewhere in Malawi.}

A different consideration in defining the outcome variable is the handling of impostors---men who come to the clinic in order to collect the voucher payment, but do not appear to be the original participant who was recruited into the study. There are only ten of these individuals in our sample; our main analysis codes them as zeroes for the outcome variable. In Appendix \autoref{tab_impostors} we code them as ones instead. This leaves the coefficient estimates nearly unchanged, and does not affect any of our inferences based on the $q$-values.
\section{Mechanisms}
\label{sec_mechanisms}

Why are appointments so effective at increasing HIV testing? Visiting a clinic for an HIV test is immediately costly in terms of time, effort and anxiety, while the benefits of treatment accrue over the longer term. This combination of short term costs with long term benefits can lead to self-control problems (\citealt{odonoghueDoingItNow1999}). Indeed, qualitative data from the baseline survey indicate that self-control problems may be an important barrier to HIV testing. When asked about reasons for avoiding an HIV test, men most commonly answer that testing is not needed, either due to low risk, a recent test, or a lack of symptoms (34 percent of participants). The next most common answer is that the participant says he is too busy, too lazy, or too forgetful to seek a test (33 percent), with 8 percent of participants mentioning laziness specifically.

Appointments can be viewed as a bundle of several behavioral interventions that help overcome self-control problems, each of which has previously been studied in isolation. This section provides evidence that two aspects of appointments play a particularly important role in their success. First, appointments appear to act as social commitment devices---providing larger self-control benefits than financial commitment devices, and avoiding their downsides. Second, appointments help overcome limited memory problems, making people less likely to forget to follow through with an HIV test. We also explore several other potential mechanisms, finding only limited evidence for them. However, our findings do not imply that commitment and limited memory are the only channels through which appointments operate.

\subsection{Appointments as Social Commitment Devices\label{section_appointments_as_commitments}}

\autoref{fig_testing_by_arm}---which is based on Column 2 of \autoref{tab_itt}---shows that the combination of appointments and commitment devices has roughly the same effect on HIV testing as the appointments-only treatment ($q$-value = 0.35). In contrast, the marginal effect of the appointments treatment on top of the commitment device treatment is positive---an increase of nine percentage points---and statistically significant at the 0.05 level. This suggests that appointments strongly substitute for commitment devices, but not vice versa. To see this, consider a scenario in which a subset of men get tested if offered only a commitment device, but do not get tested if offered only an appointment. Then, we would expect these men to get tested when offered a commitment device on top of an appointment. The fact that we do not observe an increase in the treatment effect suggests that either such a subset does not exist, or that there is a negative interaction effect---that is, adding a second intervention on top of the first actually decreases demand for HIV testing for some men. While we cannot rule out this sort of negative interaction effect \textit{a priori}, this possibility seems fairly unlikely. We therefore conclude that, for the subset of men who respond to commitment devices, appointments are just as effective. 

Additional evidence that appointments are powerful substitutes for commitment devices comes from \autoref{fig_HTE_by_CD_demand} (which is based on Column 2 of \autoref{tab_hte_by_commitment}). This figure shows that appointments are more than twice as effective for men who wanted a commitment device (compared to the effect for men who did not), and this difference is significant at the 0.1 level. The effect of the combined treatment (appointments plus commitment devices) on men who wanted a commitment device is around 23 percentage points. This is only slightly larger than the effect of the appointments-only treatment for men who wanted a commitment device, and we cannot reject the null of equal treatment effects. This comparison provides additional evidence that the appointments are very good substitutes for commitment devices: men who wanted a commitment device are almost as well-served by just getting an appointment as they are if a commitment device is layered on top of an appointment---and much better off than if they received only a commitment device.\footnote{Another piece of evidence that appointments substitute for commitment devices is that the demand for the two is positively correlated ($\rho$ = 0.2). Of the 308 men in the appointments arms who wanted a commitment device, 76 percent signed up for an appointment; for the 328 men who did not want a commitment device, just 56 percent signed up for an appointment.}

How do appointments substitute for financial commitment devices? One strong possibility is that they operate as a kind of ``social commitment'', imposing costs on the men who fail to show up. Missing an appointment means wasting the time of the HDA who was expecting you for your HIV test. Relatedly, men who miss appointments may expect their absence to be noticed and discussed by clinic staff. Similar social commitments are used as an explicit intervention by \cite{karlan_borrow_2012} to address indebtedness. They also play a role in the traditional design of microcredit products (\citealt{de_aghion_economics_2005}). 

Social pressure or social commitments work because individuals care about how they are perceived and are willing to modify their behavior to signal socially-desirable traits or to comply with social standards. The degree to which an individual's actions will be affected by social pressure can depend on how socially desirable an action is and on how much the individual cares about how they are perceived (\citealt{bursztynSocialImageEconomic2017}). This has been found to be true for behaviors ranging from vaccination (e.g., \citealt{raoSocialNetworksVaccination2007}, \citealt{karingSocialSignalingChildhood2018}) to  productivity at work (\citealt{masPeersWork2009}) to group savings (e.g., \citealt{kastUnderSaversAnonymousEvidence2012}). 
Peer effects can be stronger with proximity (e.g., \citealt{godlontonPeerEffectsLearning2012}), but proximity is not necessary (e.g., \citealt{kincaidSocialNetworksIdeation2000}). Even weak social ties can exert pressure on behavior. For example, savings deposit collectors work in part because they involve a social interaction in which it may be embarrassing or otherwise costly to disappoint the collector (\citealt{ashrafReviewCommitmentSavings2003}). The collector is able to exert social pressure despite being outside of the client's immediate social circle.

The interaction between the HDA and the participant likely exerts a form of social pressure. It may also be viewed as a social \textit{invitation} or \textit{encouragement} to visit the clinic. Clinics in Malawi are predominantly female spaces: they primarily target women through a combination of policy, practice and gender norms (\citealt{dovel2020gendered}). Offering appointments might reassure men that they will be welcomed at the clinic and provided quality care. \cite{nyondo_invitation_2015} find that men are more likely to attend antenatal care visits with their partners if they receive a formal invitation, and appointments may play a similar role. On the other hand, almost half of the men who scheduled appointments got tested on \textit{other} days, so the value of coming in on the day of the appointment is not sufficiently large to cause everyone to do so.

\subsection{Appointment Reminders and Limited Memory\label{section_appointments_as_reminders}}

While appointments are highly effective for men who want to sign up for commitment devices, they also work quite well for men who do not. These men could simply lack self-control problems, or they could be sophisticated about their self-control issues and believe they would not follow through on a commitment (\citealt{bryan2010commitment}). Either alternative presents a puzzle: why do appointments work for men who do not have self-control issues at all? Why do men who think they would not follow through on a commitment end up getting HIV tests anyway? A similar question is raised by the effects of the appointments treatment on the men who do demand commitment devices. If appointments are simply a type of commitment device, why is their failure rate so much lower? That is, why do they work more than twice as well?

Our data suggests that a second mechanism is at play in the success of appointments as well: appointment reminders help address problems of limited memory. People often forget to do things they want to do, and moreover they are overconfident about remembering their plans (\citealt{ericson_forgetting_2011}). Models of limited memory predict large effects of reminders, particularly for people who are present-biased (\citealt{ericson_interaction_2017}).\footnote{\cite{haushofer_cost_2015} shows that a model of limited memory can also generate several departures from standard neoclassical utility maximization, including status quo bias and loss aversion.}

To examine this mechanism, we plot histograms of the delay between recruitment and getting an HIV test (\autoref{figure_delays_before_testing}). Relative to both the control group (Panel A) and the commitment device arm (Panel B), men in the appointments arm got tested for HIV substantially later than those in the other two arms, and these differences are significant at the 0.01 level. The hollow red bars in the histograms show that in the non-appointments arms, testing trails off to nearly zero within about 15 days of enrollment into the study. This pattern is consistent with a limited memory mechanism, and suggests that reminders do play a role. 

\begin{landscapefig}
\begin{figure}[p!]
\caption{Histograms of Delays before HIV Test\label{figure_delays_before_testing}}
\begin{subfigure}[b]{0.6\textwidth}
\includegraphics[width=\textwidth]{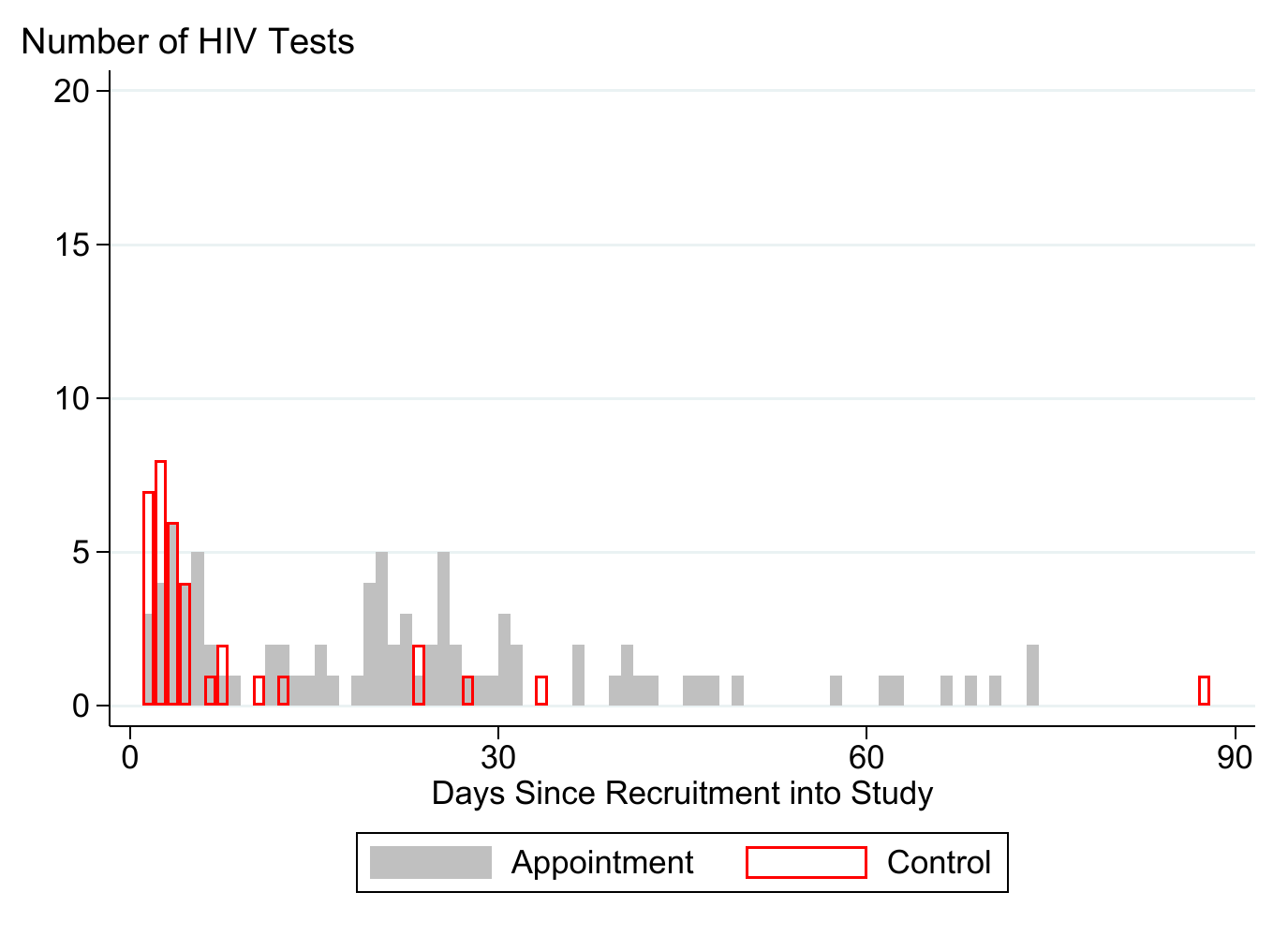}
\caption*{\textbf{Panel A:} Appointments vs. Control Group}
\end{subfigure}
\begin{subfigure}[b]{0.6\textwidth}
\includegraphics[width=\textwidth]{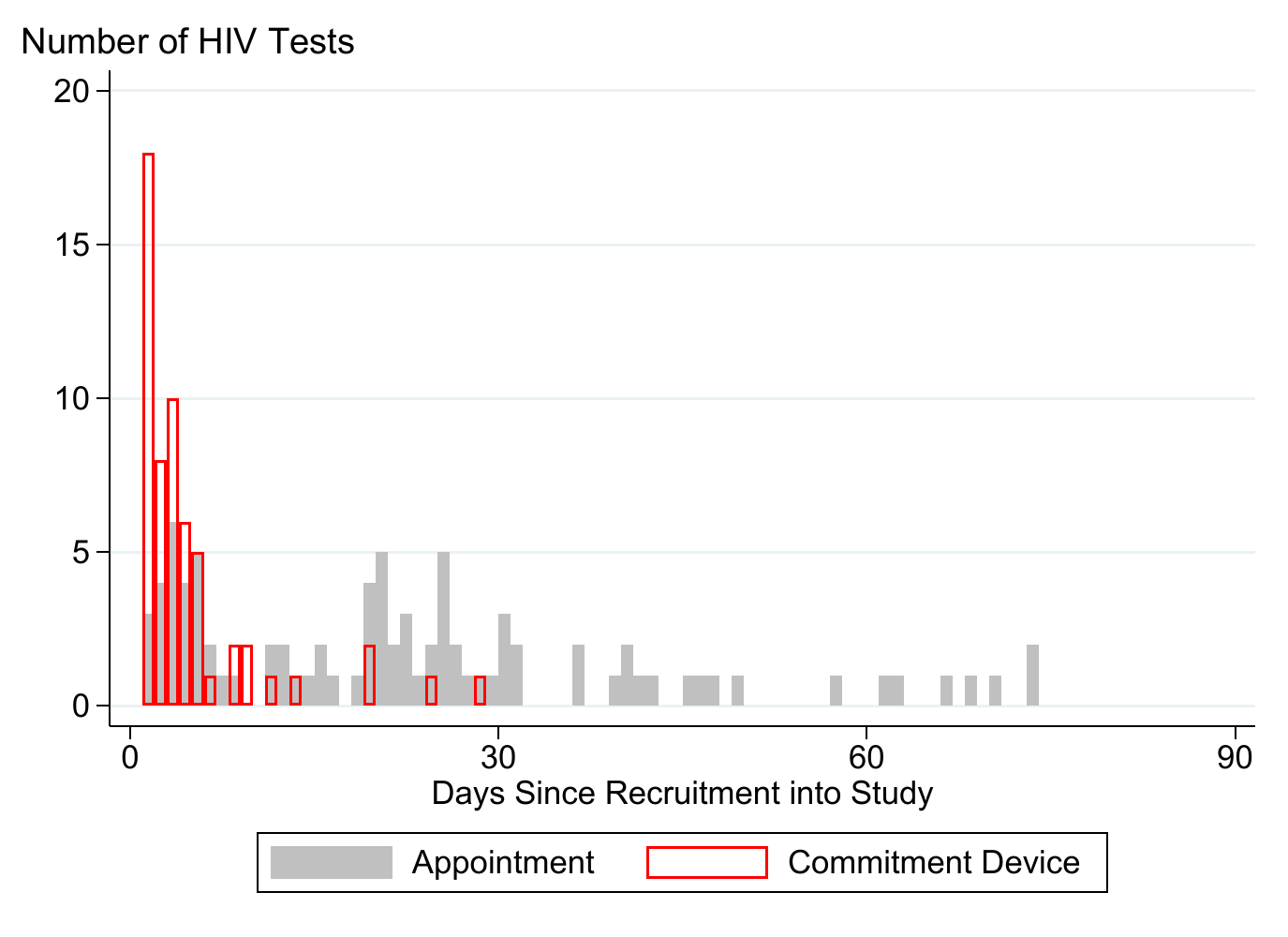}
\caption*{\textbf{Panel B:} Appointments vs. Commitment Devices}
\end{subfigure}
\begin{figurenotes}[Notes]Histogram of HIV test timing relative to the date the subject was recruited, i.e. the baseline survey date. Sample includes only men who got an HIV test, from the control group ($N=35$) the pure appointments arm ($N=87$), and the pure commitment device arm ($N=58$). $t$-tests for the equality of the average delay before an HIV test: Panel A, $p<0.001$, $q=0.001$; Panel B, $p<0.001$, $q=0.001$. 
\end{figurenotes}
\end{figure}
\end{landscapefig}
One alternative explanation for the patterns in \autoref{figure_delays_before_testing} is simply that men signed up for appointments on later dates, and followed through with them. To address this possibility, \autoref{figure_delays_by_appt_date_vs_no} breaks down the appointments vs. control comparison by whether the men in the appointments arm came in on the day of their appointment or on a different day.\footnote{Men were allowed to reschedule their appointments when they received the reminder phone call; in these cases, we use the date of the rescheduled appointment. In order to do this, we drop 64 men from the appointments arm for whom appointment rescheduling information is not available, 22 of whom took an HIV test. No men are dropped from the control group because appointment rescheduling does not apply to them. Overall testing rates are similar for men who kept their original appointment and those who rescheduled their appointments. For men who kept their original appointments, 20 percent got tested on the day of their appointment, 17 percent got tested on another day, and 63 percent did not get tested at all. For those who rescheduled, 18 percent came in on the new (rescheduled) appointment date, 17 percent came in on another date, and 65 percent did not get tested at all.} The solid gray bars are substantially shifted to the right relative to the hollow red bars in both panels, indicating that men in the appointments arm got tested later even when they did not come in on the day of their appointment. $t$-tests show that these differences are statistically significant at the 0.05 level for men who got tested on their appointment date, and at the 0.01 level for men who did not. 

The simplest explanation for the fact that men in the appointments arm test later in the study period is that the appointment reminders helped them overcome limited memory problems. The pattern in Panel B of \autoref{figure_delays_by_appt_date_vs_no} implies that a number of men in the appointments arm forgot about their plan to get tested for HIV, and then were reminded about it by the phone call about their appointment. They then followed through, but \textit{not} on their appointment date---driven to get an HIV test by the reminder, and not the social cost of a missed appointment.\footnote{An alternative explanation of this pattern is that men who miss their appointments feel rising shame from their missed appointments (\citealt{butera_measuring_2021}), inducing them to continue to come in for tests. We cannot completely rule out this potential mechanism, but it would likely lead to increased testing over time in the appointments arm, which we do not observe. It also cannot explain why the men miss their appointments in the first place.} This explanation mirrors the findings of randomized experiments that study the effects of phone call or message reminders directly, and show that that they increase attendance at medical appointments (\citealt{gurolurganci_mobile_2013}, \citealt{altmannNudgesDentist2014}). In the particular context of HIV testing, \cite{salvadori2020appointment} find that reminders explain one third of the increase in testing caused by an appointment.

\begin{landscapefig}
\begin{figure}[p!]
\caption{Histograms of Delays before HIV Test by Whether Test was on Appointment Date\label{figure_delays_by_appt_date_vs_no}}
\begin{subfigure}[b]{0.6\textwidth}
\includegraphics[width=\textwidth]{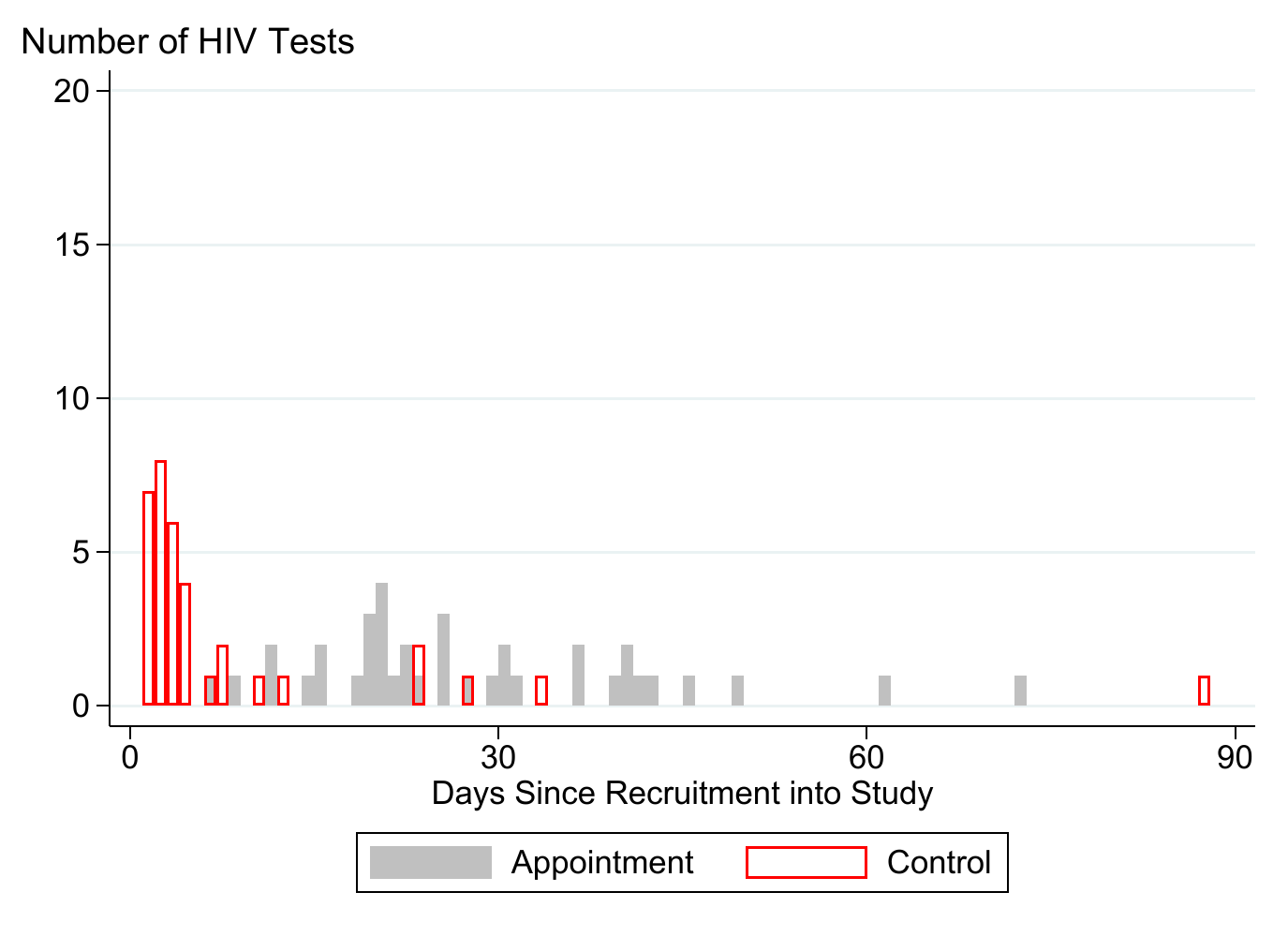}
\caption*{\textbf{Panel A:} Tested on Appointment Date}
\end{subfigure}
\begin{subfigure}[b]{0.6\textwidth}
\includegraphics[width=\textwidth]{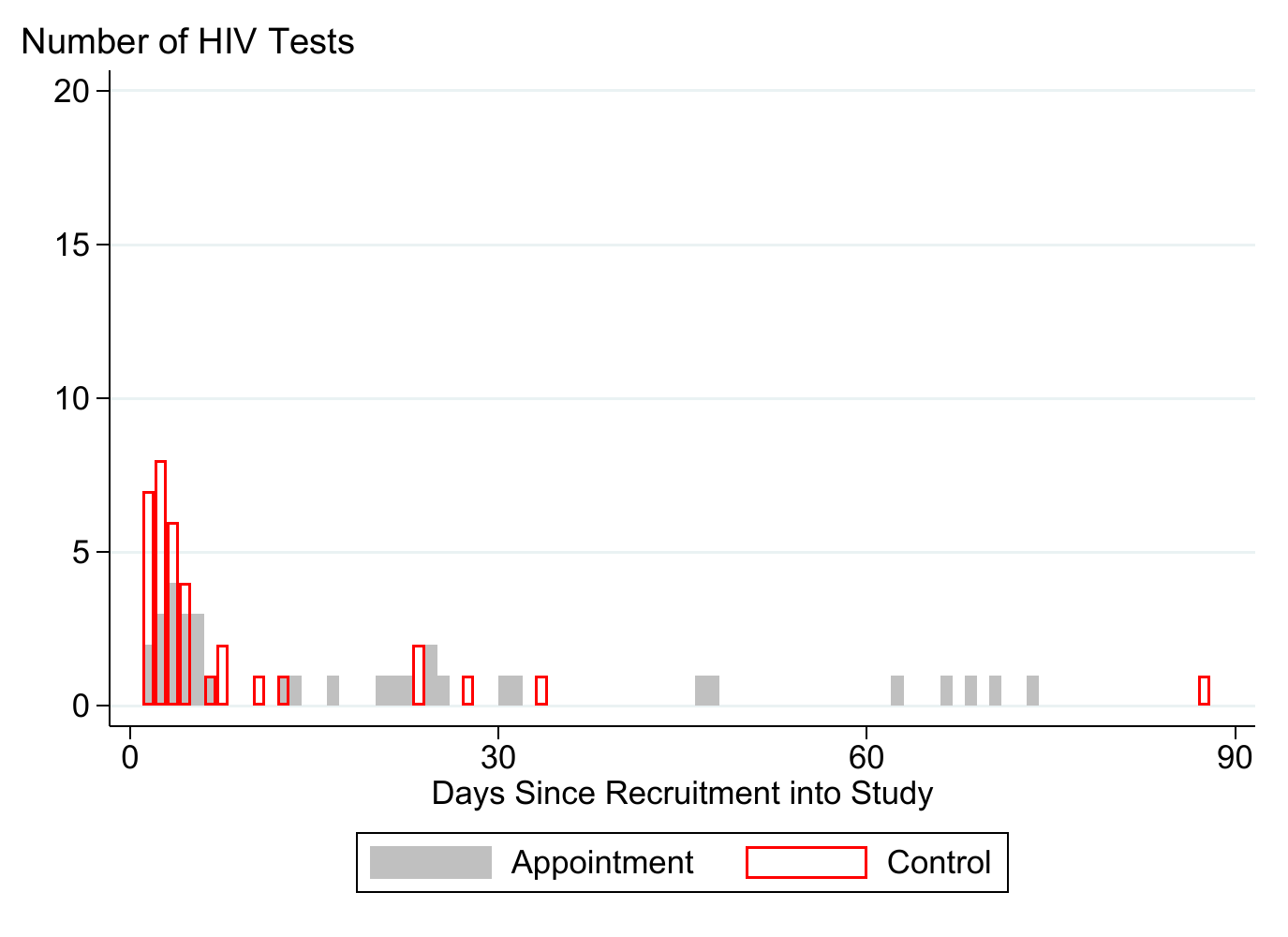}
\caption*{\textbf{Panel B:} Tested on a Another Date}
\end{subfigure}
\begin{figurenotes}[Notes]Histogram of HIV test timing relative to the date the subject was recruited, i.e. the baseline survey date. Sample includes all control-group men who got an HIV test ($N=35$), and men from the pure appointments arm who got an HIV test and for whom appointment rescheduling information is available ($N=79$). $t$-tests for the equality of the average delay before an HIV test: Panel A, $p=0.009$, $q=0.018$; Panel B, $p<0.001$, $q=0.001$. 
\end{figurenotes}
\end{figure}
\end{landscapefig}

However, appointment reminders, on their own, do not appear to fully explain our results. \autoref{figure_test_timing} shows evidence of a large spike in HIV testing on the exact appointment date: over 50 percent of all men in the two appointments arms who got an HIV test did so on the date of their appointment, and conditional on testing after the reminder, 67 percent visit the clinic exactly two days later, on the appointment date. This spike in testing is inconsistent with a pure reminder effect, which would generate an increase in visits throughout the days or weeks after the reminder. The spike is, however, consistent with a social commitment mechanism, as well as several other mechanisms explored below.

\subsection{Other Possible Mechanisms}

A different reason that appointments might increase HIV testing, and cause a spike in testing on the appointment date, is that they reduce the expected time cost of getting tested for HIV. This is one reason that appointments are common in developed-country healthcare systems: they solve coordination problems, leading to less wasted time. From the perspective of the men in our study, two related aspects of the appointments could help lower the expected time cost of HIV testing. The first is reduced wait times: people with appointments can reasonably expect to be seen close to the time of the appointment, with little to no delay. The second is the availability of providers (in this case, the HDAs who perform the HIV tests). In many developing countries, healthcare providers are routinely absent from their posts (\citealt{banerjee_addressing_2006}), so there is a strong likelihood that any trip to a clinic will simply be a waste of time. It is reasonable to expect an appointment to help address this problem, in part due to the social commitment mechanism discussed in \autoref{section_appointments_as_commitments}.\footnote{In actual fact, wait times were equalized across study arms and all clinics were staffed continuously by HDAs employed by our project. This was done to minimize the extent to which men were discouraged from completing a test by queues at the testing sites, and also to avoid overburdening the clinics with patients from our study. However, this information was not revealed to the study participants, in order to ensure that the appointments were realistic simulations of an actual appointment, including expectations about wait times and provider absenteeism.} The existence of appointments might also have made participants think the system was more efficient and competently run in general. This could have long-run benefits, but the immediate effects on HIV testing would likely operate through the provider-availability or wait-time channels.

\begin{figure}[p!]
\centering
\caption{Timing of HIV Tests Relative to Appointment Date\label{figure_test_timing}}
\includegraphics{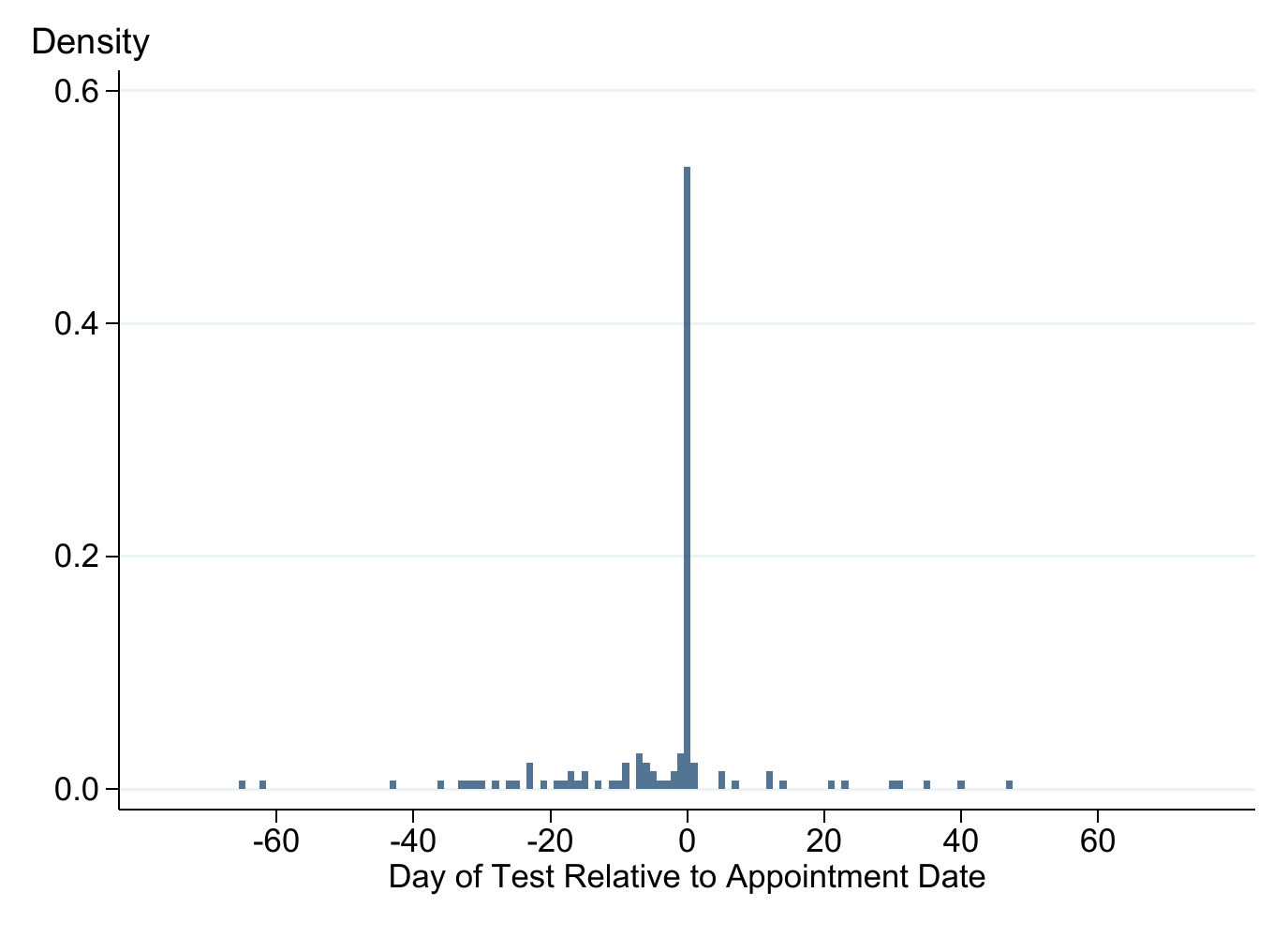}

\begin{figurenotes}[Notes]Histogram of HIV test timing relative to appointment date (using the new date for men who rescheduled their appointments). Sample includes 131 who were in one of the appointment arms and got an HIV test and for whom appointment rescheduling information is available
\end{figurenotes}
\end{figure}

We can shed further light on these possible mechanisms using qualitative interview data on typical wait times and absenteeism in 2019 for clinics in Malawi (see \ref{sec_qualwait}). We find that at the time of the study, there was typically no wait for an HIV test at any of the clinics in Zomba, though at busy times the wait could be as long as 60 minutes. Moreover, provider absence is rare: fewer than two percent of HDAs were absent on a typical day. These patterns are similar to those for Malawi as a whole. Given these facts about the clinics under the \textit{status quo}, it is highly unlikely that expected wait times and provider availability are major drivers of the effects of appointments. Indeed, in the baseline survey fewer than 0.1 percent of respondents named wait time as a reason they had avoided HIV testing.

However, several other mechanisms could explain the spike in testing on the appointment date. As discussed above, it could be that participants get tested on their exact appointment date because the HDA is expecting them, and this creates social pressure. An appointment also might make a specific date more salient, in a sense similar to \cite{bordalo_salience_2013}.\footnote{For a broader review of the literature on salience in economics, see \cite{bordalo_salience_2021}.} If the appointment date ``stands out'' in this way, we would expect a spike in attendance on the appointment date, just as we see in \autoref{figure_test_timing}. One way that raising the salience of a given date could increase testing is by reducing problems of limited memory: rather than having to come in to get an HIV test at some point, men have a specific date in mind, which is easier to remember. This would be the same fundamental mechanism as we discuss in \autoref{section_appointments_as_reminders}, but with a different empirical implication: it would predict a spike in testing on the day of the appointment. 

Finally, and somewhat related to date salience, appointments could increase testing, and increase testing on the appointment date, by prompting men to make a mental plan to get an HIV test. Prompting people to make a plan is effective at improving follow-through for many behaviors (\citealt{rogers_beyond_2015}), but planning prompts alone do not appear to work for HIV testing (\citealt{macis_using_2021}). This type of mechanism might well play a role in our setting, although the magnitude of the effect we observe is substantially larger than that typically seen for interventions that simply encourage participants to make a plan.

While not a mechanism \textit{per se}, another explanation for our results is that they are driven purely by displacement: men who would have gotten tested in the near future simply move their tests forward in time, and the intervention causes no meaningful increase in the HIV testing rate. This explanation seems unlikely for three reasons. First, the time pattern of testing in the control group shows almost no testing after the first couple of weeks. If testing were simply displaced slightly earlier in time in the appointments arm, we would expect testing to continue in the other arms, rather than stopping. Second, the time window for the study was intentionally vague, with no sharp cutoff. Most men who wanted to get tested more than two months after their recruitment into the study could still do so, and could still redeem their voucher, so we would pick up some fraction of any later HIV tests rather than missing them. Third, the annual rate of HIV testing is fairly low in Zomba. \cite{derksen_love_2021} find that on average just 5 percent of the population per year voluntarily seek out an HIV test; 26 percent of our appointments arm got an HIV test in less than three months. 

\section{Conclusion}
\label{sec_conclusion}

We show that appointments increase HIV testing more effectively than financial commitment devices. Using a randomized controlled trial in Malawi, we find that offering men appointments for HIV testing more than doubles testing rates. There is also high demand for financial commitment devices, and these traditional commitment devices significantly increase HIV testing rates. However, appointments dominate financial commitment devices, in terms of demand, treatment effects, and cost effectiveness.

Appointments also avoid an important downside of commitment devices, which is that they can be welfare-diminishing due to failed commitments (\citealt{john_when_2020,bai_self-control_2020}). None of the men in our study lost money due to missed appointments because no money was at stake.

Our results suggest that appointments work by acting as social commitment devices, creating social pressure to follow through with HIV tests. Appointments are superior substitutes for commitment devices, and have far stronger treatment effects for men who express \textit{ex ante} demand for commitment. They also address limited memory problems by providing men with reminders to come to their appointments: testing quickly trails off to zero in the control group, but continues in the appointments arm---even for men who do not actually come in on their appointment date.

While we find that these two mechanisms are important, appointments should be viewed as a coherent bundle of interventions. These interventions, many of which have been shown effective on their own, address different and overlapping behavioral biases simultaneously. The large effects we observe suggest that the whole may be greater than the sum of the parts. 

Health appointments are a natural and inexpensive policy tool, and are more feasible and straightforward to implement than a traditional commitment device. Appointments are attractive to health providers and policymakers because they allocate time and resources efficiently. Even in poor countries, mobile phone ownership is becoming increasingly common, so it is easy to schedule appointments and issue reminders. Using a commitment device for healthcare requires collecting money, or providing and withholding incentives at scale. Adding financial transactions to a healthcare system is likely to introduce managerial and logistical challenges, with no added benefit relative to simply using appointments.

Seeking an HIV test is an important yet fraught decision. The stakes are high, and access to treatment is lifesaving. The fact that offering a simple appointment can impact such an important decision is a promising sign that appointments may be effective more broadly, in healthcare settings and beyond. Studying the role of appointments in other contexts and for other demographics is an important direction for future research.

The use of appointments in developed countries---and their near absence in the developing world---may help explain the severe underutilization of healthcare in developing countries. Developing countries should seek to integrate appointments into their healthcare systems, and offer appointments for future visits to people who come in for care.

\nocite{sadoff_dynamic_2020,alan_patience_2015,halpern_randomized_2015,toussaert_revealing_2019,sadoff_can_2019,milkman_holding_2014,schwartz_healthier_2014,bhattacharya_nudges_2015}

\bibliographystyle{aea2}
\bibliography{library.bib} 
\addcontentsline{toc}{section}{References}

\pagebreak
\begin{landscape}
\end{landscape}

    \processdelayedfloats
    \makeatletter
    \efloat@restorefloats
    \makeatother

\clearpage
\appendix 

\begin{center}
\LARGE \textbf{Supplemental Online Appendix,\\Not Intended for Publication}
\end{center}

\setcounter{table}{0}
\setcounter{figure}{0}

\renewcommand{\thesection}{Appendix A}
\captionsetup*[figure]{name=Appendix Figure}
\renewcommand\thefigure{A\arabic{figure}} 
\captionsetup*[table]{name=Appendix Table}
\renewcommand{\thetable}{A\arabic{table}}

\section{Qualitative Data\label{sec_qual}}

In this section we summarize qualitative data collected from staff at clinics in Zomba and across Malawi. We also include a selection of direct quotes that illustrate our key findings. The full, deidentified qualitative dataset will be made available upon request. 

\subsection{Appointments for HIV Testing\label{sec_qualappthiv}}

In general, appointments are not used for HIV testing. There are some exceptions, in some clinics, for people that are referred for HIV testing from the outpatient department (OPD) and patients that were treated for a sexually-transmitted infection (STI). Clinics also sometimes contact the sexual partners of people who have tested positive and invite them to get tested on a particular day. More generally, in the rare cases where ``appointments’’ are scheduled for HIV testing, they simply involve inviting a person to come in for a test within a certain number of weeks, or sometimes on a particular day, without scheduling a specific time or using formal reminders.

\begin{itemize}
\item[] ``[When asked if they schedule appointments for HIV testing]: There were some who would do that but it was not common. These would be the people who are very busy like the business people they would call and tell us they want to come for testing on a particular day and time. We also have other clients who would tell us that they can only come to the clinic on Saturday because of the nature of their work they cannot come during the week.’’ -- HDA at a large public clinic
\item[] ``We have appointments for the AIT [Active Index Testing]\footnote{``Active Index Testing'' refers to the practice of offering HIV tests to all sexual contacts identified by a person recently diagnosed with HIV.} clients because we have already made arrangements that they will come to the clinic on a day that we have booked them. […] We also have the STI clients because we give them an appointment to come after four weeks.’’ -- HDA at a large public clinic
\item[] ``We don’t specify the time when the client should come we just tell them come on Wednesday or Thursday [...] [On the appointment day] we first see the person who booked an appointment, but we inform the people on the queue that this client had an appointment and we have to see him first because we already started the process in the past and we would want to complete the process.'' -- HDA at a large public clinic
\end{itemize}

\subsection{Appointments for Other Services\label{sec_qualapptother}}

Appointments are also rare in clinics more generally. They are not typically used for OPD services. Patients receiving treatment for HIV are told to come back around a certain date or whenever they need a prescription refill.  Clinics do report using informal appointments for various services as needed, but most patients are walk-in, and wait times are short. At large hospitals, highly specialized services, such as surgery and oncology visits, are scheduled for a particular date. But, even in this case patients are not given a particular time, nor formal reminders.

\begin{itemize}
\item[] ``It is the nurses and the clinicians in the wards who assess the clients and for those that need medical attention are taken to OPD so there is no need for appointment. For the ART we know and we keep their medication and we give	them at the time they are supposed to take the medication; this time we are giving them in the morning. And when the drugs are coming to an end we go and refill at the ART clinic so we don’t have to make any appointments.'' -- Staff at a small public clinic
\item[] ``[When asked if they use appointments]: If we have the surgical clinic and other clinics; for the surgical clinic they are done on Mondays. [...] We write these clients in the appointment books; we have hard covers which we enter the names of the clients. [...] For this specific clinic we do not give them specific time because we have a room where the clinic is conducted and it is done on first come and first served and the clinician can also make consideration if there are clients who are very sick and needed to be seen first. The clinic runs from 7:30 am and ends at 4:30 pm and people are advised to come within this time.'' -- Staff at a large public hospital
\item[] ``[When making appointments]: We give them the specific date but we don’t give them the specific time. [...] To be honest we do not give them that chance [to choose the appointment date] [...] we just do the booking from our perspectives not considering the client’s choice.'' -- Staff at a large public clinic
\item[] ``For cough, flu and malaria do not require appointments and we have the chronic conditions these require appointments; like those which needs further investigations and treatment they require appointments. [Of those who visit the clinic] 10 percent of the clients [have] appointments.'' -- Staff at a small public clinic
\end{itemize}

\subsection{HIV Testing Wait Times and Absenteeism\label{sec_qualwait}}

In July 2019, during the study period, every clinic in our sample consistently had excess capacity for HIV testing. In small clinics, there were multiple HDAs on duty, yet on the average day fewer than ten clients came for HIV testing. Larger clinics saw as many as 12 patients per HDA. HIV testing takes as little as 5 minutes (if the result is negative) or as long as 45 minutes (with counselling). So, it appears that even large clinics were operating below capacity. HDAs report that there was typically no wait for HIV testing, but at busy times, at large clinics, wait times could be as long as one hour. Clinics are well-staffed, and absenteeism among HDAs is extremely rare.

\begin{itemize}
\item[] ``It was 4 on average. […] 3 HDAs work in the clinic.’’ -- HDA at a small public clinic
\item[] ``We would have 36 clients on a daily basis. We open every day except for the weekend. […] We were two HDAs now we have a new one.'' -- HDA at a large public clinic
\item[] ``The patients would wait between 15 to 20 minutes because the patient flow was very fast.’’ -- HDA at a large public clinic
\item[] ``The client will wait for his turn between 40 to 60 minutes. But if there is no client in the counselling room it means he will enter immediately.'' -- HDA at a large public hospital
\item[] ``I don’t think there was anyone who was absent. People [HDAs] work up to the knocking off time.’’ -- Senior Staff at a large public clinic
\end{itemize}

\clearpage

\setcounter{table}{0}
\setcounter{figure}{0}

\renewcommand{\thesection}{Appendix B}
\captionsetup*[figure]{name=Appendix Figure}
\renewcommand\thefigure{B\arabic{figure}} 
\captionsetup*[table]{name=Appendix Table}
\renewcommand{\thetable}{B\arabic{table}}

\section{Cost Calculations\label{sec_costs}}

This section details how we calculated the costs of each intervention. We compute only the incremental cost of each pure treatment arm: the \textit{additional} labor costs and other expenses needed to carry out the intervention, beyond the costs that apply to all study arms including the control group. Our cost calculations thus exclude the cost of finding and contacting men for the survey, since that was done in the control group as well. The labor costs are primarily calculated based on differences in the total survey time. To compute these, we only use survey times for respondents who were assigned either into the control group or into one of the two pure intervention arms (that is, we omit the combined intervention arm). 

\subsection{Appointments Arm}

The cost of offering an appointment has two main components: (i) the cost of explaining and scheduling an appointment during the baseline survey, and (ii) the cost of reminding the participant of his appointment and rescheduling if necessary.

Part (i) is the product of enumerator wages and the average time it takes to explain and schedule an appointment, taking into account the fact that some participants who are offered an appointment do not schedule one. This time difference has two elements: (a) the time explaining appointments and eliciting demand, and (b) the time spent adding the chosen time slot into the calendar for those who want an appointment. We can measure (a) by calculating the time the SurveyCTO application was active for each participant and taking the difference between the average time in the appointments-only group and the average time in the control group. Part (b) was done in a different application on the phone, so it is not included in the active on-screen time on the survey app. We estimate this time to be one minute per person, based on having our research team carry out the steps needed to schedule the appointment and timing themselves. We therefore add one minute to the times of those participants who actually scheduled an appointment. The total time that corresponds to  part (i) equals 139.4 seconds. Enumerators were paid MK20,000 for 6 hour days, so MK3,333.33 per hour. The cost of part (i) is therefore MK129. The following equation shows the formula used to compute part (i) of the cost of the appointments:

\begin{equation*}
   \text{Part (i)}^A = \big( \text{Extra Survey Time} + \mathbbm{1}[ \text{Scheduled Appt.}] \times (60 \text{ seconds}) \big) \times \frac{\text{Enumerator Hourly Wage}} { 3,600 \text{ seconds per hour}}   
\end{equation*}

Part (ii) has two components: wages paid to HDAs who make the calls and phone credit spent on successful calls. Times per call were not estimated during the experiment. We asked our research team to estimate these times by calling people and reading them the appointment reminder script and timing themselves. We found that it could take up to 40 seconds for a respondent to pick up, and we estimated the time it takes to read the reminder script to be 65 seconds. Wages must be paid for this entire time (105 seconds) but phone credit is only charged for active call time. Every person who schedules an appointment receives a reminder call. Out of those calls, 42 percent of men who initially scheduled an appointment rescheduled their appointment once, and thus received a second reminder for the rescheduled appointment. At that reminder, 16 percent of  men who initially scheduled an appointment rescheduled for a second time, and thus received a third reminder. We account for rescheduling times differently than reminder times. Reminder calls took 65 seconds each, and some of those included rescheduling time (estimated to be 40 seconds). 

To price the time used on the calls we need to know the wage rate for the HDAs who made the calls, and the cost of phone credit. HDAs were paid MK20,000 for 8 hour days. There are different telecommunication companies in Malawi, and they charge different rates per minute of calls. In our study we used the two main companies that operate in the area. One of them charges MK65 per minute and the other charges MK72. In these calculations we take a conservative approach by using the higher of the two rates to calculate phone credit costs. This brings the cost of part (ii) to be MK185 The following equation depicts the formula used to compute part (ii) of the cost of the appointments.

\begin{align*}
    \begin{split}
    \text{Part (ii)}^A =   & \biggl(\frac{\text{  Calls Made}}{\text{Appointments Offered}}  \times \text{ Reminder Script Time} 
   \\ &+ \frac{\text{Rescheduled Appointments}}{\text{Appointments Offered}}  \times \text{ Rescheduling Time} \biggr) \\ &\times \left( \frac{\text{Credit cost/minute}}{60 \text{ seconds/hour}} + \frac{\text{HDA Hourly Wage}}{3,600 \text{ seconds/hour}} \right)
    \\
    + &\frac{\text{ Calls made}}{\text{Appointments Offered}}  \times \text{ Waiting Time} \times  \frac{\text{HDA Hourly Wage}}{3,600 \text{ seconds per hour}}
    \end{split}
\end{align*}

The total cost of adding a person to the appointment arm is the sum of parts (i) and (ii), which comes to MK314.07 
 
\subsection{Commitment Device Arm}

The cost of offering a commitment device has two parts: (i) the extra survey time needed to explain and enroll men in the commitment device, and (ii) the cost of the voucher used for the commitment device. Since we elicited demand for the commitment device from all of our participants, average survey time differences between those assigned only a commitment device and those in the pure control group (16 seconds) do not  measure part (i) completely. In order to estimate it, we asked our field team to read the survey script that explains the intervention and time themselves. We find that this part of the survey takes on average 180 seconds. Therefore we add 180 seconds to the average survey time differences between pure control and those only offered a commitment device. These time differences are then multiplied by the wages of enumerators, yielding a total cost for part (i) of MK181.46.  The following equation shows the formula used to compute part (i) of the cost of the commitment devices.

\begin{equation*}
    \text{Part (i)}^{CD} = \left( \text{Extra Survey Time} + \text{CD Explanation Script Time} \right) \times \frac{\text{Enumerator Hourly Wage}}{3,600 \text{ seconds per hour}} 
\end{equation*}

Computing the correct cost to include for part (ii) is less straightforward. If the commitment device is implemented with the participant's own money, the total cost of adding a participant to the commitment device arm would be simply part (i). If the program were to use its own funds, then the amount to be committed needs to be budgeted for. It may be necessary to budget for the  MK1,000 for all participants, bringing the total cost to MK1,181.46. But not all participants who accept a commitment device will follow through and collect the money. In our study, 49 percent of men in the commitment device-only study arm enrolled in a commitment device, but just 41 percent of those men redeemed their voucher. Considering the fact that the project kept the MK1,000 for the 29 percent of men who did not redeem their voucher, the net cost of offering the vouchers is just MK710.9 per person. Thus the cost of enrolling one man in the commitment device arm would come to MK892.36.
\clearpage

\setcounter{table}{0}
\setcounter{figure}{0}

\renewcommand{\thesection}{Appendix C}
\captionsetup*[figure]{name=Appendix Figure}
\renewcommand\thefigure{C\arabic{figure}}  
\captionsetup*[table]{name=Appendix Table}
\renewcommand{\thetable}{C\arabic{table}}

\section{Appendix Tables\label{sec_appendix_tables}}

\begin{landscape}
\begin{table}[htbp!]

\caption{Pre-Specified Control Variables and Definitions  \label{tab_controls_prespec}}

\includegraphics[width=\linewidth,page=6]{Tables/Appointments_Tables_cropped.pdf}
\begin{tablenotes}[Notes]Pre-specified list of control variables from our analysis plan. These controls were selected as statistically significant predictors of past HIV testing choices, based on the baseline data from the study. 
\end{tablenotes}

\end{table}
\end{landscape}

\begin{landscape}
\begin{table}[ht!]

\caption{Other Control Variables and Definitions  \label{tab_controls_other}}

\includegraphics[width=\linewidth,page=7]{Tables/Appointments_Tables_cropped.pdf}
\begin{tablenotes}[Notes]Other variables used in our balance table and in the double LASSO procedure. This list includes all other baseline variables that were not included in the pre-specified list in \autoref{tab_controls_prespec}
\end{tablenotes}
\vspace{-60pt}
\end{table}
\end{landscape}

\begin{table}[!htbp]
\caption{Balance\label{tab_balance}} \includegraphics[width=\linewidth,page=8]{Tables/Appointments_Tables_cropped.pdf}
\end{table}

\begin{table}[!htbp]
\ContinuedFloat
\caption{Balance (continued)}    
\includegraphics[width=\linewidth,page=9]{Tables/Appointments_Tables_cropped.pdf}
\begin{tablenotes}[Notes]Sample is 1,232 men who completed a baseline survey. Panel A compares all men assigned to either appointments or appointments + commitment devices (treatment) to everyone else (control); Panel B compares all men assigned to either commitment devices or appointments + commitment devices (treatment) to everyone else (control). Columns 3 and 8 present treatment-control differences, with $p$-values from $t$-tests in parentheses: * $p<0.1$; ** $p<0.05$; *** $p<0.01$. Columns 4 and 9 show differences in standard deviations. Columns 5 and 10 present OLS regressions of the treatment indicators on all the baseline covariates in the table; the bottom row of the table shows $F$-statistics (and associated $p$-values) for the joint significance of all the variables in the regression.
\end{tablenotes}
\end{table}

\begin{table}[htbp!]
\caption{Effects of Appointments and Commitment Devices on HIV Testing---Short Specification\label{tab_itt_short}} 
\includegraphics[width=\linewidth,page=10]{Tables/Appointments_Tables_cropped.pdf}
\begin{tablenotes}[Notes]Sample is 1,232 men who completed a baseline survey. Pre-specified controls include all the variables in \autoref{tab_controls_prespec}. Double Lasso controls uses the \cite{chernozhukov2017double} method for variable selection and inference, as described in \autoref{sec_itt}. Columns 2 and 3 also control for date-of-survey fixed effects, enumerator fixed effects, and preferred clinic fixed effects. Heteroskedasticity-robust standard errors in parentheses: * $p<0.1$; ** $p<0.05$; *** $p<0.01$; \cite{anderson2008multiple} sharpened $q$-values in brackets.
\end{tablenotes}
\vspace{-10pt}
\end{table}

\begin{table}[htbp!]
\centering
\caption{Treatment Effect Heterogeneity by Demand for Commitment---Short Specification\label{tab_hte_by_commitment_short}} 
\includegraphics[width=\linewidth,page=11]{Tables/Appointments_Tables_cropped.pdf}
\begin{tablenotes}[Notes]Sample is 1,232 men who completed a baseline survey. These regressions do not include a constant; (CD) x (1-D) is omitted and assumed to be zero. Pre-specified controls include all the variables in \autoref{tab_controls_prespec}. Double Lasso controls uses the \cite{chernozhukov2017double} method for variable selection and inference, as described in \autoref{sec_itt}. Columns 2 and 3 also control for date-of-survey fixed effects, enumerator fixed effects, and preferred clinic fixed effects. All controls are standardized prior to running the regressions. Heteroskedasticity-robust standard errors in parentheses: * $p<0.1$; ** $p<0.05$; *** $p<0.01$; \cite{anderson2008multiple} sharpened $q$-values in brackets.
\end{tablenotes}
\vspace{-80pt}
\end{table}

\begin{table}[htbp!]
\caption{Treatment Effect Heterogeneity by Baseline Covariates\label{tab_hte}} 
\includegraphics[width=\linewidth,page=12]{Tables/Appointments_Tables_cropped.pdf}
\begin{tablenotes}[Notes]Sample is 1,232 men who completed a baseline survey. Pre-specified controls include all the variables in \autoref{tab_controls_prespec}. Other baseline covariates interacted with treatments in Column 5 include all the other variables from \autoref{tab_controls_prespec}. Main effects are included for all variables that are interacted with the treatment indicators. We also control for $A \times CD$ and its interactions with all the baseline variables in the table, but do not show these results. Heteroskedasticity-robust standard errors in parentheses: * $p<0.1$; ** $p<0.05$; *** $p<0.01$; \cite{anderson2008multiple} sharpened $q$-values in brackets.
\end{tablenotes}
\end{table}

\begin{table}[htbp!]
\caption{Treatment Effect Heterogeneity by Predicted Probability of Getting Tested\label{tab_estrat}} 
\includegraphics[width=\linewidth,page=13]{Tables/Appointments_Tables_cropped.pdf}
\begin{tablenotes}[Notes]Sample is 1,232 men who completed a baseline survey; Columns 1 and 2 use only data from the control (C) and appointments-only (A) arms, while Columns 3 and 4 use only data from the control and commitment devices-only (CD) arms. This table uses the repeat split-sample method of \cite{abadie2018endogenous} to predict the outcome variable in the control group, then estimates treatment effects separately by tercile of the predicted outcome. Pre-specified predictors include all the variables in \autoref{tab_controls_prespec}. Double Lasso predictors include the same variables chosen using the \cite{chernozhukov2017double} in \autoref{tab_itt}. All columns control for date-of-survey fixed effects, enumerator fixed effects, and preferred clinic fixed effects. Heteroskedasticity-robust standard errors in parentheses: * $p<0.1$; ** $p<0.05$; *** $p<0.01$; \cite{anderson2008multiple} sharpened $q$-values in brackets. $p$-values for the tests of equality between terciles ignore the covariance term and thus are likely to be conservative.
\end{tablenotes}
\vspace{-30pt}
\end{table}

\begin{table}[htbp!]
\caption{Effects on Positive HIV Tests and ART Initiation\label{tab_hivpos_art}} 
\includegraphics[width=\linewidth,page=14]{Tables/Appointments_Tables_cropped.pdf}
    \begin{tablenotes}[Notes]Sample is 1,232 men who completed a baseline survey. The regressions in this table include no controls or fixed effects because HIV status and ART initiation data was anonymized and linked only to participants' study arms. Heteroskedasticity-robust standard errors in parentheses: * $p<0.1$; ** $p<0.05$; *** $p<0.01$; \cite{anderson2008multiple} sharpened $q$-values in brackets.
    \end{tablenotes}
\end{table}

\begin{table}[htbp!]
\caption{Randomization Inference\label{tab_ri}} 
\includegraphics[page=15]{Tables/Appointments_Tables_cropped.pdf}
\begin{tablenotes}[Notes]Sample is 1,232 men who completed a baseline survey. This table presents the same results as Columns 4-6 of \autoref{tab_itt}, but shows randomization inference (RI) $p$-values for comparison with the $p$-values that correspond to the Eicker–Huber–White (EHW) standard errors that we show in the rest of the tables in the paper. Pre-specified controls include all the variables in \autoref{tab_controls_prespec}. Double Lasso controls uses the \cite{chernozhukov2017double} method for variable selection and inference, as described in \autoref{sec_itt}. Columns 2 and 3 both control for date-of-survey fixed effects, enumerator fixed effects, and preferred clinic fixed effects. Heteroskedasticity-robust standard errors in parentheses: * $p<0.1$; ** $p<0.05$; *** $p<0.01$; \cite{anderson2008multiple} sharpened $q$-values in brackets.
\end{tablenotes}
\end{table}

\begin{landscape}
\begin{table}[htbp!]
\caption{Effects on Voucher Redemption\label{tab_voucher}} 
\includegraphics[width=\linewidth,page=16]{Tables/Appointments_Tables_cropped.pdf}
\begin{tablenotes}[Notes]Sample is 1,232 men who completed a baseline survey. Pre-specified controls include all the variables in \autoref{tab_controls_prespec}. Double Lasso controls uses the \cite{chernozhukov2017double} method for variable selection and inference, as described in \autoref{sec_itt}. Columns 2, 3, 5, and 6 also control both control for date-of-survey fixed effects, enumerator fixed effects, and preferred clinic fixed effects. Heteroskedasticity-robust standard errors in parentheses: * $p<0.1$; ** $p<0.05$; *** $p<0.01$; \cite{anderson2008multiple} sharpened $q$-values in brackets.
\end{tablenotes}
\end{table}
\end{landscape}    

\begin{table}[htbp!]
\caption{Robustness to Including Impostors\label{tab_impostors}} 
\includegraphics[page=17]{Tables/Appointments_Tables_cropped.pdf}
\begin{tablenotes}[Notes]Sample is 1,232 men who completed a baseline survey. This table presents the same results as \autoref{tab_itt}, but codes impostors as having a one for the outcome variable instead of a zero. Pre-specified controls include all the variables in \autoref{tab_controls_prespec}. Double Lasso controls uses the \cite{chernozhukov2017double} method for variable selection and inference, as described in \autoref{sec_itt}. Columns 2 and 3 both control for date-of-survey fixed effects, enumerator fixed effects, and preferred clinic fixed effects. Heteroskedasticity-robust standard errors in parentheses: * $p<0.1$; ** $p<0.05$; *** $p<0.01$; \cite{anderson2008multiple} sharpened $q$-values in brackets.
\end{tablenotes}
\end{table}

\clearpage

\setcounter{table}{0}
\setcounter{figure}{0}

\renewcommand{\thesection}{Appendix D}
\captionsetup*[figure]{name=Appendix Figure}
\renewcommand\thefigure{D\arabic{figure}} 
\captionsetup*[table]{name=Appendix Table}
\renewcommand{\thetable}{D\arabic{table}}

\section{Results Using Specification from Analysis Plan\label{sec_short_specification}}

This section presents the results of our analyses using the regression specification that we pre-specified in our analysis plan (\citealt{derksen_soft_2019},  \url{https://www.socialscienceregistry.org/versions/57507/docs/version/document}). This differs from the version we present in the body of the paper in that it imposes the ``short'' model, which omits the interaction term. We present all the tables that include treatment effect estimates because the $q$-values are computed across all the tables, and thus even tables which show the exact same specification are affected by the change.

\begin{landscape}
\begin{table}[htbp!]
\caption{Effects of Appointments and Commitment Devices on HIV Testing\label{tab_itt_old}} 
\includegraphics[width=\linewidth,page=1]{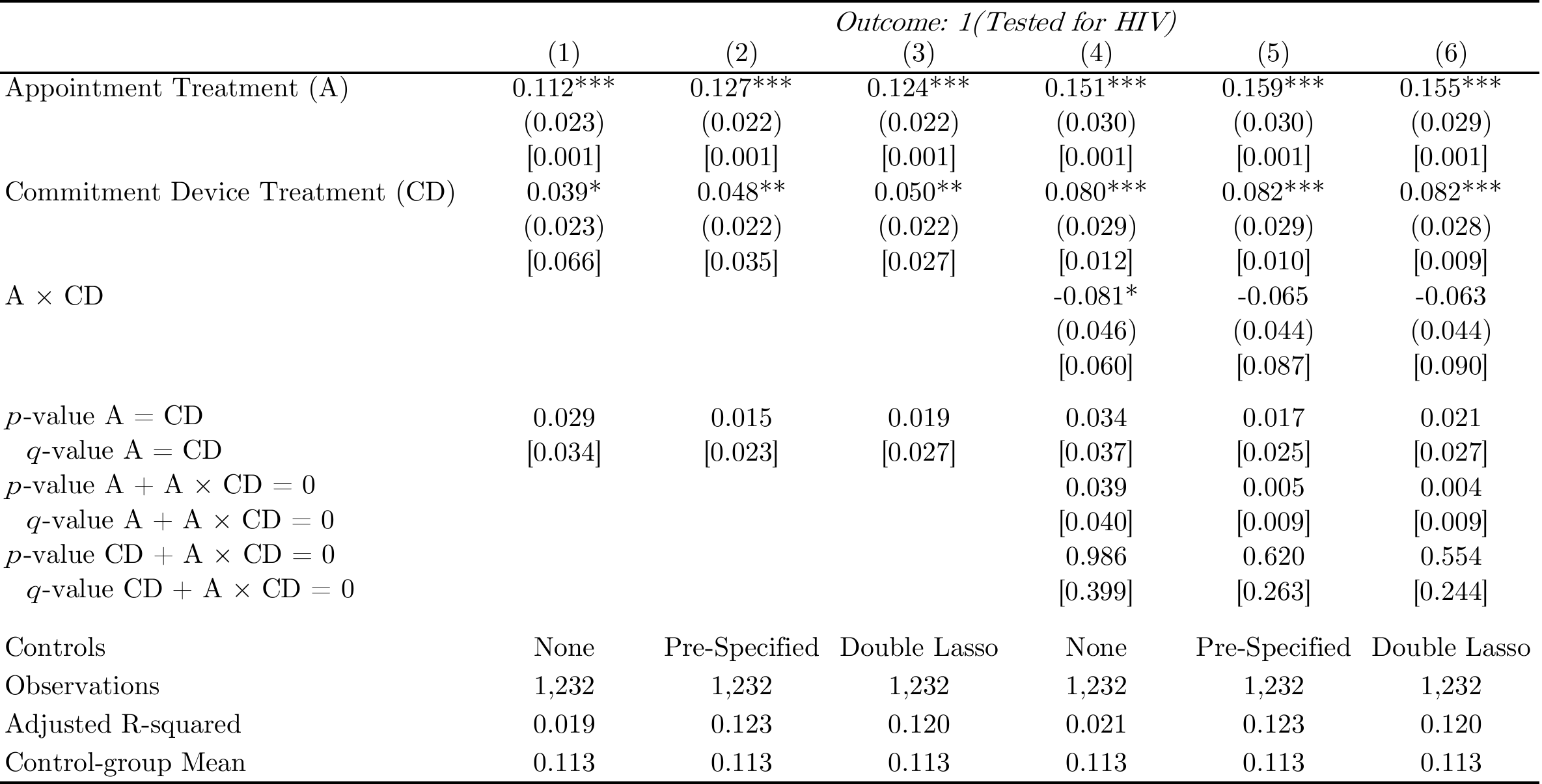}
\begin{tablenotes}[Notes]Sample is 1,232 men who completed a baseline survey. Pre-specified controls include all the variables in \autoref{tab_controls_prespec}. Double Lasso controls uses the \cite{chernozhukov2017double} method for variable selection and inference, as described in \autoref{sec_itt}. Columns 2 and 3 both control for date-of-survey fixed effects, enumerator fixed effects, and preferred clinic fixed effects. Heteroskedasticity-robust standard errors in parentheses: * $p<0.1$; ** $p<0.05$; *** $p<0.01$; \cite{anderson2008multiple} sharpened $q$-values in brackets.
\end{tablenotes}
\end{table}
\end{landscape}   

\begin{table}[htbp!]
\caption{2SLS Estimates of the Treatment-on-the-Treated Effect of Appointments\label{tab_2sls_old}} 
\includegraphics[page=2]{Tables/Appointments_Tables_old_cropped.pdf}
\begin{tablenotes}[Notes]Sample is 1,232 men who completed a baseline survey. Pre-specified controls include all the variables in \autoref{tab_controls_prespec}. Double Lasso controls uses the \cite{chernozhukov2017double} method for variable selection and inference, as described in \autoref{sec_itt}. Columns 2 and 3 both control for date-of-survey fixed effects, enumerator fixed effects, and preferred clinic fixed effects. In lieu of a conventional $F$-statistics, Panel B shows the effective $F$-statistic of \cite{montiel_olea_robust_2013}. Heteroskedasticity-robust standard errors in parentheses: * $p<0.1$; ** $p<0.05$; *** $p<0.01$; \cite{anderson2008multiple} sharpened $q$-values in brackets.
\end{tablenotes}
\end{table}

\begin{landscape}
\begin{table}[htbp!]
\caption{Effects on Voucher Redemption\label{tab_voucher_old}} 
\includegraphics[width=\linewidth,page=11]{Tables/Appointments_Tables_old_cropped.pdf}
\begin{tablenotes}[Notes]Sample is 1,232 men who completed a baseline survey. Pre-specified controls include all the variables in \autoref{tab_controls_prespec}. Double Lasso controls uses the \cite{chernozhukov2017double} method for variable selection and inference, as described in \autoref{sec_itt}. Columns 2 and 3 both control for date-of-survey fixed effects, enumerator fixed effects, and preferred clinic fixed effects. Heteroskedasticity-robust standard errors in parentheses: * $p<0.1$; ** $p<0.05$; *** $p<0.01$; \cite{anderson2008multiple} sharpened $q$-values in brackets.
\end{tablenotes}
\end{table}
\end{landscape}    

\begin{landscape}
\begin{table}[htbp!]
\caption{Robustness to Including Impostors\label{tab_impostors_old}} 
\includegraphics[width=\linewidth,page=12]{Tables/Appointments_Tables_old_cropped.pdf}
\begin{tablenotes}[Notes]Sample is 1,232 men who completed a baseline survey. This table presents the same results as \autoref{tab_itt}, but codes impostors as having a one for the outcome variable instead of a zero. Pre-specified controls include all the variables in \autoref{tab_controls_prespec}. Double Lasso controls uses the \cite{chernozhukov2017double} method for variable selection and inference, as described in \autoref{sec_itt}. Columns 2 and 3 both control for date-of-survey fixed effects, enumerator fixed effects, and preferred clinic fixed effects. Heteroskedasticity-robust standard errors in parentheses: * $p<0.1$; ** $p<0.05$; *** $p<0.01$; \cite{anderson2008multiple} sharpened $q$-values in brackets.
\end{tablenotes}
\end{table}
\end{landscape}    

\begin{table}[htbp!]
\caption{Effects on Positive HIV Tests and ART Initiation\label{tab_hivpos_art_old}} 
\includegraphics[page=13]{Tables/Appointments_Tables_old_cropped.pdf}
\begin{tablenotes}[Notes]Sample is 1,232 men who completed a baseline survey. The regressions in this table include no controls or fixed effects because HIV status and ART initiation data was anonymized and linked only to participants' study arms. Heteroskedasticity-robust standard errors in parentheses: * $p<0.1$; ** $p<0.05$; *** $p<0.01$; \cite{anderson2008multiple} sharpened $q$-values in brackets.
\end{tablenotes}
\end{table}

\begin{table}[htbp!]
\caption{Treatment Effect Heterogeneity by Baseline Covariates\label{tab_hte_old}} 
\includegraphics[width=\linewidth,page=14]{Tables/Appointments_Tables_old_cropped.pdf}
\begin{tablenotes}[Notes]Sample is 1,232 men who completed a baseline survey. Pre-specified controls include all the variables in \autoref{tab_controls_prespec}. Other baseline covariates interacted with treatments in Column 5 include all the other variables from \autoref{tab_controls_prespec}. Main effects are included for all variables that are interacted with the treatment indicators. Heteroskedasticity-robust standard errors in parentheses: * $p<0.1$; ** $p<0.05$; *** $p<0.01$; \cite{anderson2008multiple} sharpened $q$-values in brackets.
\end{tablenotes}
\end{table}

\begin{landscape}
\begin{table}[htbp!]
\caption{Treatment Effect Heterogeneity by Demand for Commitment\label{tab_hte_by_commitment_old}} 
\includegraphics[width=0.90\linewidth,page=3]{Tables/Appointments_Tables_old_cropped.pdf}
\begin{tablenotes}[Notes]Sample is 1,232 men who completed a baseline survey. Pre-specified controls include all the variables in \autoref{tab_controls_prespec}. Double Lasso controls uses the \cite{chernozhukov2017double} method for variable selection and inference, as described in \autoref{sec_itt}. Columns 2 and 3 both control for date-of-survey fixed effects, enumerator fixed effects, and preferred clinic fixed effects. Heteroskedasticity-robust standard errors in parentheses: * $p<0.1$; ** $p<0.05$; *** $p<0.01$; \cite{anderson2008multiple} sharpened $q$-values in brackets.
\end{tablenotes}
\vspace{-60pt}
\end{table}
\end{landscape}

\begin{table}[htbp!]
\caption{Treatment Effect Heterogeneity by Predicted Probability of Getting Tested\label{tab_estrat_old}} 
\includegraphics[width=\linewidth,page=4]{Tables/Appointments_Tables_old_cropped.pdf}
\begin{tablenotes}[Notes]Sample is 1,232 men who completed a baseline survey. This table uses the repeat split-sample method of \cite{abadie2018endogenous} to predict the outcome variable in the control group, then estimates treatment effects separately by tercile of the predicted outcome. Pre-specified predictors include all the variables in \autoref{tab_controls_prespec}. Double Lasso predictors include the same variables chosen using the \cite{chernozhukov2017double} in \autoref{tab_itt}. All columns control for date-of-survey fixed effects, enumerator fixed effects, and preferred clinic fixed effects. Heteroskedasticity-robust standard errors in parentheses: * $p<0.1$; ** $p<0.05$; *** $p<0.01$; \cite{anderson2008multiple} sharpened $q$-values in brackets. $p$-values for the tests of equality between terciles ignore the covariance term and thus are likely to be conservative.
\end{tablenotes}
\end{table}

\end{document}